\documentclass[12pt]{iopart}
\pdfoutput=1
\usepackage{iopams}
\usepackage{color,graphicx}% Include figure files
\usepackage{bm}% bold math
\usepackage[utf8]{inputenc}
\usepackage[pdftex]{hyperref}
\DeclareGraphicsExtensions{.pdf,.png,.jpg,.mps,.eps}

\newcommand{\bra}[1]{\left\langle {#1}\right|}
\newcommand{\ket}[1]{\left|{#1}\right\rangle}
\newcommand{\average}[1]{\left<{#1}\right>}
\renewcommand{\Re}{\mathop{\mathrm{Re}}}
\renewcommand{\Im}{\mathop{\mathrm{Im}}}
\newcommand{\B}{\mathrm{B}}
\newcommand{\C}{\mathrm{c}}
\newcommand{\MF}{\mathrm{MF}}
\newcommand{\eq}{\mathrm{eq}}
\newcommand{\st}{\mathrm{st}}
\renewcommand{\P}{\mathrm{P}}
\newcommand{\eff}{\mathrm{eff}}
\newcommand{\Br}{\mathrm{Br}}
\newcommand{\sys}{\mathrm{S}}

\newcommand{\loc}{\mathrm{loc}}

\newcommand{\Pb}[2]{\left[{#1},{#2}\right]_{\mathrm{PB}}}
\newcommand{\Order}[1]{\mathrm{O}\left({#1}\right)}
\begin{document}
% Journal identifier can be put here if required, e.g.r
%\jl{14}

\title[Entropy Production in Quantum Brownian Motion]{Entropy Production in Quantum Brownian Motion}

\author{Lorenzo Pucci\dag\footnote[2]{To whom correspondence should be addressed.}, 
Massimiliano Esposito\S\ and Luca Peliti\dag$\|$}

\address{\dag\ Dipartimento di Fisica, Università ``Federico~II'', Complesso Monte S.~Angelo, I--80126 Napoli (Italy)}

\address{\S\ Complex Systems and Statistical Mechanics,
University of Luxembourg,
Campus Limpertsberg,
162a avenue de la Faïencerie,
L--1511 Luxembourg (G. D. Luxembourg)
}

\address{$\|$ Associato Istituto Nazionale di Fisica Nucleare, Sezione di Napoli}

\eads{\mailto{pucci@na.infn.it},
\mailto{massimiliano.esposito@uni.lu},
\mailto{peliti@na.infn.it}}

\begin{abstract}
We investigate how to coherently define entropy production for a process of transient relaxation in the Quantum Brownian Motion model for the harmonic potential. We compare a form, called ``Poised" (P), which after non-Markovian transients corresponds to a definition of heat as the change in the system  Hamiltonian of mean force, with a recent proposal by Esposito \etal (ELB) based on a definition of heat as the energy change in the bath. Both expressions yield a positive-defined entropy production and coincide for vanishing coupling strength, but their difference is proved to be always positive (after non-Markovian transients disappear) and to grow as the coupling strength increases. In the classical over-damped limit the ``Poised" entropy production converges to the entropy production used in stochastic thermodynamics. We also investigate the effects of the system size, and of the ensuing Poincaré recurrences, and how the classical limit is approached. We close by discussing the strong-coupling limit, in which the ideal canonical equilibrium of the bath is violated.
\end{abstract}

\submitto{JSTAT}

%%%%%%%%%%%%%%%%%%%%%%%%%%%%%%%%%%%%%%%%%%%%%%%%%%%%%%%%%%%%%%%%%%%%%%
\section{Introduction}\label{intro} 

The theory of stochastic thermodynamics provides a consistent description of nonequilibrium thermodynamics for classical systems weakly coupled to their environments and described by Markovian dynamics \cite{Esposito12, Seifert12Rev}. In recent years fundamental characteristics of classical thermodynamics have been put under scrutiny in the quantum realm where a proper formulation of nonequilibrium thermodynamics seems a much harder task. In particular when considering low temperatures and non-vanishing couplings various difficulties arise, some of which are already present at equilibrium. 
An ubiquitous exactly solvable model to address these questions is the Quantum Brownian Motion (QBM) model \cite{Ullersma196627,Haake,Caldeira}. It consists of a system with Hamiltonian $H_\sys$ (often an harmonic oscillator) bi-linearly coupled via a term denoted $H_I$ to a bath of harmonic oscillators with Hamiltonian $H_B$. The total Hamiltonian is thus of the form $H=H_\sys+H_B+H_I$.
In this model, when the total system is initially in canonical equilibrium $\rho^\eq = \rme^{-\beta H}/Z$, the Clausius formulation of the second law seems to be violated for a quasi-static change of the mass or of the frequency of the central oscillator \cite{Allahverdyan00, HorhammerButtnerJSP08}. In this case, the heat flow is defined as the change in the averaged central system Hamiltonian due to the bath and is found to be larger than the temperature times the change in the system entropy, defined as the von Neumann entropy of the central system. To remain consistent with these definitions, work and free energy are also defined in terms of the system Hamiltonian and as a result the Thompson formulation of the second law is also violated \cite{FordOConnellPRL06, Nieuwenhuizen}. More work can be extracted from the system than the change in its free energy. 
One also intriguingly finds that the behavior of the heat capacity of the system is different when it is derived from the energy of the central system at equilibrium or from a partition function approach~\cite{HanggiIngoldTalkner08NJP, HanggiIngoldActaPol06}. In this latter case the heat capacity might even become negative at low temperature. Also, the von Neumann entropy of the central oscillator does not vanish at zero temperature while the equilibrium entropy of the total system (which coincides with the von Neumann entropy of the total system) does. These phenomena can be ascribed to quantum correlations between the central system and the bath \cite{LutzPRA09}. 
Various attempts have been made in order to overcome these difficulties. Some of them incorporate in Thompson's formulation of the second law the work contribution required to initially couple the system to the bath at zero \cite{FordOConnellPRL06} or arbitrary temperatures \cite{Hilt, KimMahler07}. Others introduce different notions of effective temperature \cite{Nieuwenhuizen, KimMahlerPRE10}. Ultimately, many of the difficulties are related to the fact that the equilibrium density matrix of the central oscillator is not the familiar canonical distribution $\rho^\eq_\sys = \rme^{-\beta H_\sys}/Z_\sys$ defined in terms of the central system Hamiltonian as is often the case in weak-coupling theories. 

In this paper, we want to investigate the slightly different problem of transient relaxation to equilibrium in the QBM model. This means that we initially place the central oscillator in a nonequilibrium state and put it in contact with its bath at equilibrium. Due to the interaction, the two parts of the system will exchange energy and if the bath is sufficiently large, the central oscillator will asymptotically reach an equilibrium state. For such a process we would like to identify a meaningful notion of entropy production. 
In stochastic thermodynamics the nonequilibrium version of the second law states that for such a relaxation process the entropy production is equal to the change in system entropy (which is identified with the Shannon entropy of the system) minus the heat exchanged with the bath divided by the bath temperature. Furthermore the entropy production can be proved to be an always positive quantity which only vanishes at equilibrium. 
For quantum systems the Shannon entropy is replaced by the von Neumann entropy of the system $S=-\tr_\sys \rho_\sys \ln \rho_\sys$ and a very similar formulation holds as long as the quantum system is weakly coupled to its bath and described by a Markovian quantum master equation \cite{Spohn1978,Breuer}. The heat exchanged with the bath is then expressed in terms of the system Hamiltonian by integrating $\dot{Q} \equiv \tr_\sys H_\sys \dot{\rho}_\sys$. An attempt to use such an expression for the entropy production for the QBM model was made in~\cite{Nieuwenhuizen} but with non satisfactory results, since a negative entropy production rate was obtained for non-vanishing system-bath coupling strength. A more satisfactory definition of entropy production has been recently introduced by Esposito, Lindenberg and Van den~Broeck~\cite{Esposito} (denoted here by ELB) in the form $\Delta_{\rm i} S=\Delta S-Q/T$, where the heat is now defined as minus the energy change in the bath ($\dot{Q} \equiv - \tr H_B \dot{\rho}_B$). This quantum entropy production is positive definite even for finite bath sizes, notwithstanding recurrences. This definition applies on the assumption that the central system and the bath have uncorrelated density matrices at the initial time. 

In this paper, we compare this definition of entropy production $\Delta_{\rmi} S$ with a new definition $\Delta_{\rm i} S^{\P}$, inspired by the one introduced in~\cite{Breuer} in the context of Markovian master equations. The corresponding heat definition is expressed in term of the averaged change of an effective Hamiltonian which reduces to the system Hamiltonian $H_S$ in the weak-coupling limit. We evaluate analytically both expressions of the entropy production for the QBM model and evaluate their difference. We consider only Gaussian initial conditions, both for the bath and for the central oscillator, what guarantees that the density matrix remains Gaussian at all times. We find that $\Delta_{\rmi} S$ is positive definite but can present oscillations while $\Delta_{\rmi} S^{\P}$ is positive definite and has a positive time derivative only in the Markovian high-temperature or weak-coupling limits. The difference between the two definitions considerably depends on the coupling.
We also study the behavior of the entropy production for finite-size thermal baths, where Poincaré recurrences characterize the time evolution of the system. The convergence towards a continuous relaxing behavior is studied as a function of the system size. It turns out that a Lorentzian, rather than uniform, sampling of oscillation frequencies of the bath guarantees a better convergence. 
Finally the evolution the von~Neumann entropy of the bath is studied. It is found that, for fixed initial conditions, its asymptotic value does not depend on the coupling in the classical limit while it does in the quantum regime. However, in both cases the Kullback-Leibler divergence between the density matrix of the bath at time $t$ and at canonical equilibrium depends considerably on the coupling, making the usual approximation of the ideal bath problematic.

\subsection*{Outline}
In section~\ref{def2law} the different definitions of the entropy production are spelled out, along with the general protocol adopted.
In section~\ref{model} the Quantum Brownian Motion model is introduced and solved. Initial conditions are specified in ~\ref{ic} and the evolution of the system is described in section~\ref{ME} via its Wigner quasi-distribution function. The approach to the thermodynamic limit is described in section~\ref{therm_limit}.
Explicit expressions of the definitions of entropy for our model are reported in section~\ref{Results}. Section~\ref{Poi_rec} is devoted to a study of the model with a finite-sized bath, where Poincaré recurrences characterize its behavior. A discussion of the bath entropy, correlation entropy and of the distance of the bath density operator from its canonical form is reported in section~\ref{bath_entropy}. In section~\ref{Conclusions} we conclude and summarize our results. A few technical details are relegated in several appendices.

%%%%%%%%%%%%%%%%%%%%%%%%%%%%%%%%%%%%%%%%%%%%%%%%%%%%%%%%%%%%%%%%%%%%%%

%%%%%%%%%%%%%%%%%%%%%%%%%%%%%%%%%%%%%%%%%%%%%%%%%%%%%%%%%%%%%%%%%%%%%%
\section{Entropy production} \label{def2law}
%%%%%%%%%%%%%%%%%%%%%%%%%%%%%%%%%%%%%%%%%%%%%%%%%%%%%%%%%%%%%%%%%%%%%%

We consider a central system $\sys$ coupled to its bath $\B$. The total Hamiltonian is:
\begin{equation}
 H = H_\sys+H_\B+H_\mathrm{I}.
\end{equation}
We assume to prepare the system and the bath separately, so that no correlation is initially present between them, and to instantaneously switch on the interaction $H_\mathrm{I}$ at $t=0$. We also assume that the bath is initially at canonical equilibrium. The density matrix of the total system is therefore of the form
\begin{equation}\label{factor_ic}
\rho(0) = \rho_\sys(0) \otimes \rho_\B(0) \ \ , \ \
\rho_{\B}(0)=\rho_{\B}^{\eq} \equiv \frac{\rme^{-\beta H_{\B}}}{Z_{\B}}\ \ , \ \ 
Z_{\B}=\tr_{\B} \rme^{-\beta H_{\B}},
\end{equation}
where $\beta=(k_\mathrm{B}T)^{-1}$ is the Boltzmann factor and $\rho_\sys(0)$ and $\rho_{\B}(0)$ are respectively the central oscillator and the bath reduced density matrix. From now on we set $k_{\mathrm{B}}=1$.
The density matrix $\rho_{\sys}(t)$ of the system $\sys$ evolves according to the equation
\begin{equation}\label{st-nonloc}
 \rho_\sys(t)=V(t)\rho_\sys(0)=\tr_\B \rho(t),
\end{equation}
with the evolved total density operator 
\begin{eqnarray}\label{rho_tot_ev}
\rho(t) = U(t) \rho(0) U^\dagger(t),
\end{eqnarray} 
where
\begin{eqnarray}
U(t) =\rme^{-\rmi Ht},\label{eq:unitary}
\end{eqnarray} 
is the unitary evolution operator in the global system $\mathrm{S}\otimes\mathrm{R}$.

The evolution of the system density matrix (\ref{st-nonloc}) can be formally written as 
\begin {equation}
 \dot{\rho_\sys}(t)=\mathcal{L}(t)\rho_\sys(t) \ \ , \ \ \mathcal{L}(t) = \dot{V}(t)V^{-1}(t) , \label{MEq}
\end {equation}
where the operator $\mathcal{L}(t)$ in general depends on time. In this case, the evolution operator $V(t)$ may be written as $V(t)=\mathcal{T} \exp{\{\int_0^t d\tau \mathcal{L}(\tau)\}}$ in which $\mathcal{T}$ indicates the time-ordering operator. The form (\ref{MEq}) is the starting point to derive convolutionless quantum master equations \cite{Breuer}. In the Markovian case, $\mathcal{L}(t)$ is time-independent. This typically happens in the thermodynamic limit of the bath for times $t$ larger then the bath correlation time.

%--------------------------------------------------------------
\subsection{The `Poised' definition}\label{BreuerSect}

Let us define the \textit{poised} density matrix $\rho^{*}_{\sys}(t)$ as the solution of
\begin{equation}
 V(t) \rho^{*}_{\sys}(t) = \rho^{*}_{\sys}(t).
\end{equation}
Its existence and uniqueness may not always be guaranteed. In the Markovian case, the poised density matrix $\rho^{*}_{\sys}$ becomes constant in time and coincides with the stationary density matrix $\rho^{\st}_{\sys}$, defined by $\mathcal{L} \rho^{\st}_{\sys}=0$:
\begin{equation}
\rho^{*}_{\sys} = \rho^{\st}_{\sys}. \label{AllTheSame}
\end{equation}
For the QBM model, the poised density matrix is well defined at all times. Its expression is given in equation (\ref{ln_poised}) and is derived in~\ref{app:breuer}.

We can then introduce the following definition of the entropy production: 
\begin{equation}\label{DS_i_poi}
 \Delta_{\rm i} S^{\P}= \left[D(\rho_\sys(0)\|\rho^{*}_{\sys}(t))-D(\rho_\sys(t)\|\rho^{*}_{\sys}(t))\right] ,
\end{equation}
where $D(\ldots\|\ldots)$ is the Kullback-Leibler divergence, defined by
\begin{equation}\label{eq:KL}
D(\rho\|\rho')=\tr \rho \ln \rho -\tr \rho \ln \rho' \geq 0.
\end{equation}
One can prove that the expression (\ref{DS_i_poi}) is positive definite as follows. 
From equation (\ref{st-nonloc}) we obtain
\begin{eqnarray} 
D(\rho_\sys(t)\|\rho^{*}_{\sys}(t))
&=&D(V(t)\rho_{\sys}(0)\|V(t)\rho_{\sys}^{*}(t)) \nonumber\\
&=&D\left(\tr_{\B}U(t)\rho_{\sys}(0)\otimes\rho_{\B}(0)U^{\dag}(t)\|\tr_{\B}U(t)\rho^{*}_{\sys}(t)\otimes\rho_{\B}(0)U^{\dag}(t)\right)\nonumber\\
&\le& D(\rho_{\sys}(0)\|\rho^{*}_{\sys}(t)), \label{Inequal}
\end{eqnarray}
where we have used the property of the Kullback-Leibler divergence
\begin{equation}
 D(\rho_1\|\rho_2) \geq D(\tr_\B\rho_1\|\tr_\B\rho_2)  .
\end{equation}

We also introduce the slightly different entropy production
\begin{equation}\label{DS_i_eq}
 \Delta_{\rm i} S^{\Br}=D(\rho_\sys(0)\|\rho_\sys^{\st})-D(\rho_\sys(t)\|\rho_\sys^{\st}) ,
\end{equation}
which is also obviously also positive definite in the Markovian case and extends the definition previously proposed by Breuer and Petruccione for weakly coupled systems \cite{Breuer} where $\rho_\sys^{\st}$ is the canonical distribution expressed in terms of the system Hamiltonian $H_\sys$.

In the Markovian case, it follows from (\ref{Inequal}) that the time derivative of $\Delta_{\rm i} S^{\P}$ as well as that of $\Delta_{\rm i} S^{\Br}$ is also positive \cite{Breuer,Yukawa}:
\begin{equation}
\frac{d \Delta_{\rm i} S^{\P}}{dt} = \frac{d \Delta_{\rm i} S^{\Br}}{dt} 
=-\lim_{dt \to 0} \frac{D(\rho_\sys(t+dt)\| \rho^{\st}_{\sys})-D(\rho_\sys(t)\| \rho^{\st}_{\sys})}{dt} \ge 0.
\end{equation}
This result does not hold in the non-Markovian case where $\mathcal{L}(t)$ depends on $t$ and where $d \Delta_{\rm i} S^{\P}/dt \neq d \Delta_{\rm i} S^{\Br}/dt$. 

It may happen that the stationary density matrix can be expressed as the canonical distribution of some effective Hamiltonian $H_\sys^{\eq}$. In the weak-coupling limit it corresponds to the system Hamiltonian $H_\sys$. Its expression for the QBM model is given in equation~(\ref{H_S_eq}). In such cases, the entropy production (\ref{DS_i_eq}) has a straightforward physical interpretation, since it can be rewritten as
\begin{equation} \label{DS_i_eqbis}
 \Delta_{\rm i} S^{\Br}=\Delta S-\Delta_{\rm e} S^{\Br},
\end{equation}
i.e., as the difference between the entropy change $\Delta S$, and the entropy flow $\Delta_{\rm e} S^{\Br}$ identified by the variation of the averaged effective Hamiltonian of the central system:
\begin{equation}\label{DS_e_eq}
 \Delta_{\rme} S^{\Br}=\beta\Delta\average{H_\sys^{\eq}}.
\end{equation}
The average of an operator $O$ is defined as $\average{O}=\Tr\rho(t) O$ and $\Delta$ denotes the difference between the average evaluated at time $t$ and at time $0$.
We shall see in section~\ref{Mark_Kramers} that in the classical high-temperature limit and in the quantum weak-coupling limit this definition becomes equal to the one used in the usual stochastic thermodynamics setup. In the following, when studying the QBM model for a finite-size bath, where $\rho^{\st}_{\sys}$ does not exist, and when referring to $\Delta_{\rm i} S^{\Br}$, one should consider the definition (\ref{DS_i_eqbis}) instead of (\ref{DS_i_eq}). When the bath approaches the thermodynamic limit the two definitions are equivalent.

%--------------------------------------------------------------
\subsection{The ELB definition}

A definition of the entropy production which guarantees its positivity in all cases has been recently introduced by Esposito \etal~\cite{Esposito}. It reads
\begin{equation}\label{DS_i_esp}
 \Delta_{\rm i} S = D\left(\rho(t)\big\|\rho_\sys(t)\otimes \rho^{\eq}_{\B}\right).
\end{equation}
This quantity is clearly positive definite and within our assumptions it expresses the second law in the form
\begin{equation}\label{DS_esp}
 \Delta_{\rm i} S = \Delta S + \beta\,\Delta\average{H_\B}.
\end{equation}
One has in fact, exploiting the conservation of the von~Neumann entropy:
\begin{eqnarray}
\hspace{-2.5cm}
D\left(\rho(t)\|\rho_{\sys}(t)\otimes \rho_{\B}^{\eq}\right)&=&\tr \rho(t)\ln \rho(t)-\tr \rho(t)\ln \rho_{\sys}(t)-\tr \rho(t)\ln \rho_{\B}^{\eq}\nonumber\\
&=&\tr \rho_{\B}^{\eq}\ln \rho_{\B}^{\eq}+\tr \rho_{\sys}(0)\ln \rho_{\sys}(0)-\tr \rho_{\sys}(t)\ln \rho_{\sys}(t)-\tr \rho_{\B}(t)\ln \rho_{\B}^{\eq}\nonumber\\
&=&\Delta S(t)+\left[\tr \rho_{\B}^{\eq}\left(-\beta H_{\B}\right)-\tr \rho_{\B}(t)\left(-\beta H_{\B}\right)\right].
\end{eqnarray}
From (\ref{DS_esp}) we observe that the entropy flow
\begin{equation}\label{DS_e_esp}
\Delta_{\rm e} S=-\beta\, \Delta\average{H_\B}=\beta \big( \Delta \average{H_\sys} + \Delta \average{H_\mathrm{I}} \big),
\end{equation}
is now identified as the change of the bath energy $H_\B$ times the inverse temperature $\beta$, as usual in equilibrium thermodynamics. 
It is then proportional to the change of the central oscillator energy plus an interaction term which is discussed in section~\ref{Results}.
The time derivative of (\ref{DS_i_esp}) is not generally positive, as we will observe in \ref{Res-CL}. Since by (\ref{DS_e_eq}) and (\ref{DS_e_esp}) one has $H^{\eq}\rightarrow H_\sys$ and $\average{H_\mathrm{I}}\rightarrow0$, we see that the two definitions coincide in the weak-coupling limit.
We note that the positivity of the right-hand-side of (\ref{DS_esp}) was also remarked in ref.~\cite{Jarzynski99}.

%--------------------------------------------------------------
\subsection{Difference between the two definitions}\label{diff-DSi}

Since the following identity holds
\begin{equation}
 D(\rho(t)||\rho_\sys^{\st}\otimes\rho_\B^{\eq})=D(\rho(t)||\rho_\sys(t)\otimes\rho_\B^{\eq})+D(\rho_\sys(t)||\rho_\sys^{\st}),
\end{equation}
we find, using  (\ref{DS_i_esp}) and (\ref{DS_i_eq}), that the difference between the ELB definition $\Delta_{\rm i} S$ and the Breuer one $\Delta_{\rm i} S^{\Br}$ is given by
\begin{equation}\label{DSi-DSiBr}
 \Delta_{\rm i} S-\Delta_{\rm i} S^{\Br}=D(\rho(t)||\rho_\sys^{\st}\otimes\rho_\B^{\eq})-D(\rho_\sys(0)||\rho_\sys^{\st}).
\end{equation}
Due to the unitary evolution of the total density operator $\rho(t)$ (\ref{rho_tot_ev},\ref{eq:unitary}), one can recast the first term of the right hand side of (\ref{DSi-DSiBr}) in the form
\begin{equation} 
 D(\rho(t)||\rho_\sys^{\st}\otimes\rho_\B^{\eq})=D(\rho_\sys(0)\otimes\rho_\B^{\eq}||U^\dagger(t)\rho_\sys^{\st}\otimes\rho_\B^{\eq}U(t)).
\end{equation}
By inserting this identity into (\ref{DSi-DSiBr}) and noting that
\begin{equation}
 D(\rho_\sys(0)||\rho_\sys^{\st})=D(\rho_\sys(0)\otimes\rho_\B^{\eq}||\rho_\sys^{\st}\otimes\rho_\B^{\eq}),
\end{equation}
we obtain
\begin{eqnarray}\label{diffEspBr}
  \Delta_{\rm i} S-\Delta_{\rm i} S^{\Br}
=-\tr \rho_\sys(0)\otimes\rho_\B^{\eq}[\ln U^\dagger(t)\rho_\sys^{\st}\otimes\rho_\B^{\eq}U(t)-\ln \rho_\sys^{\st}\otimes\rho_\B^{\eq}].
\end{eqnarray}
Moreover due to the inequality
\begin{eqnarray}
  &&\hspace{0cm}D(\rho_\sys(0)\otimes\rho_\B^{\eq}||U^\dagger(t)\rho_\sys^{\st}\otimes\rho_\B^{\eq}U(t))\nonumber\\
  &&\qquad\qquad\hspace{-1cm}\geq D(\tr_\B\{\rho_\sys(0)\otimes\rho_\B^{\eq}\}||\tr_\B\{U^\dagger(t)\rho_\sys^{\st}\otimes\rho_\B^{\eq}U(t)\})\nonumber\\
  &&\qquad\qquad\hspace{-1cm}=D(\rho_\sys(0)||\tr_\B\{U^\dagger(t)\rho_\sys^{\st}\otimes\rho_\B^{\eq}U(t)\}),
\end{eqnarray}
it follows from (\ref{diffEspBr}) that
\begin{eqnarray}\label{DSi-DSiBrdis}
 \Delta_{\rm i} S-\Delta_{\rm i} S^{\Br}\geq -\tr_\sys \rho_\sys(0) \left[\ln \tilde{V}(t)\rho_\sys^{\st}-\ln \rho_\sys^{\st}\right],
\end{eqnarray}
where we have introduced the evolution operator for the central system associated to the total adjoint dynamics, implicitly defined by
\begin{equation}
 \tilde{V}(t)\rho_\sys=\tr_\B\{U^\dagger(t)\rho_\sys\otimes\rho_\B^{\eq}U(t)\},
\end{equation}
where $\rho_\sys$ is a generic density operator for the system $\sys$.

If the operator $\mathcal{L}(t)$ is Markovian, $\tilde{\mathcal{L}}(t)=\dot{\tilde{V}}(t)\tilde{V}^{-1}(t) $ will also be so. If they furthermore have the same stationary state, so that
\begin{equation}\label{stat-V_VTR}
 \tilde{V}(t)\rho_\sys^{\st}=\rho_\sys^{\st},
\end{equation}
then the right-hand side of (\ref{DSi-DSiBrdis}) vanishes and the ELB expression is strictly larger than the Breuer one:
\begin{equation}\label{Poi>Bre}
 \Delta_{\rm i} S-\Delta_{\rm i} S^{\Br} \geq 0.
\end{equation}
In~\ref{sec:TR_dynamics} we show that this is indeed the case in the QBM model. 

In the following we are going to study these different definitions in the context of the QBM model \cite{Ullersma196627,Haake,Caldeira}. 

%%%%%%%%%%%%%%%%%%%%%%%%%%%%%%%%%%%%%%%%%%%%%%%%%%%%%%%%%%%%%%%%%%%%%%%%%%%%%%%%%%%%
\section{The model}\label{model}
%%%%%%%%%%%%%%%%%%%%%%%%%%%%%%%%%%%%%%%%%%%%%%%%%%%%%%%%%%%%%%%%%%%%%%%%%%%%%%%%%%%%

The QBM Hamiltonian represents an harmonic oscillator bi-linearly 
coupled with coupling constants $\epsilon_i$ to a bath of $N$ harmonic oscillators:
\begin{equation}\label{Hamilt}
\hspace{-1cm} H_\sys =\frac{1}{2}\left(\omega_0^2Q_{0}^2+P_{0}^2\right) \ \ , \ \ 
H_\B =\frac{1}{2}\sum_{i=1}^N \left(\omega_i^2Q_i^2+P_i^2\right) \ \ , \ \ 
H_\mathrm{I} =\sum_{i=1}^N \epsilon_iQ_{0}Q_i. \nonumber
\end{equation}
We have put all masses equal to one for simplicity.

The equations of motion in the Heisenberg picture read
\begin{eqnarray}\label{H-equations}
\dot{Q_\mu}(t)=\frac{\rmi}{\hbar}\left[H,Q_\mu(t)\right] \ \ , \ \ \dot{P_\mu}(t)=\frac{\rmi}{\hbar}\left[H,P_\mu(t)\right].
\end{eqnarray}
The Greek indexes $\mu$ and $\nu$ include by convention also the central oscillator and the terms associated to the bath, while the Latin ones only run on the bath degrees of freedom. For convenience, we will use the shorthand notation $Q=Q_0$ and $P=P_0$. The solution of the equations of motion reads~\cite{Ullersma196627,Haake}
\begin{eqnarray}\label{H-solutions}
Q_\mu(t) =\sum_{\mu=0}^N\left(\dot{A}_{\mu\nu}(t)Q_\nu(0)+A_{\mu\nu}(t)P_\nu(0)\right) \ \ , \ \ P_\mu(t) =\dot{Q}_\mu(t).
\end{eqnarray}
 The form assumed by the $A_{\mu\nu}(t)$'s is reported for completeness in (\ref{Usol}).

One also gets a condition to be fulfilled in order to obtain a positive-definite Hamiltonian and non diverging solutions (see (\ref{g_z})):
\begin{equation}\label{eq:omega0}
\Omega_0^2=\omega_0^2-\sum_{i=1}^N\frac{\epsilon_i^2}{\omega_i^2}\ge 0.
\end{equation}
This expression actually defines a normalized frequency of the central oscillator, $\Omega_0$, as it appears in the Quantum Langevin Equation (QLE) picture \cite{Fleming}(cf.~\ref{QLE}).

The term $A(t)\equiv A_{00}$ plays the role of a retarded propagator. This can be seen by putting the solutions (\ref{H-solutions}) for the central
oscillator in the form \cite{Fleming}
\begin{equation}\label{Lang_sol}
{\bm z}(t)={\bm \Phi}(t){\bm z}(0)-({\bm \Phi}\ast{\bm \eta})(t),
\end{equation}
where we have defined $\bm{z}(t)$ with ${\bm z}^{\mathsf{T}}(t)=({Q}(t), {P}(t))$, the matrix propagator
\begin{equation}\label{M_prop}
{\bm \Phi}(t)=\left[\begin{array}{c c}
\dot{A}(t) & A(t)\\
\ddot{A}(t) & \dot{A}(t)
\end{array}\right],
\end{equation}
and the noise $\bm{\eta}^{\mathsf{T}}(t)=(0, {\eta}(t))$ with components
\begin{equation}\label{eta}
{\eta}(t)=\sum_i^N\epsilon_i\left[Q_i(0)\cos\omega_it+\frac{P_i(0)}{\omega_i}\sin\omega_it\right].
\end{equation}
This is actually the solution of the QLE reported in~\ref{QLE}), from which it appears that the dynamics is characterized by a damping kernel
\begin{equation}
K(t)  = \int_0^\infty \rmd\omega\;\frac{\gamma(\omega)}{\omega^2}\,\cos\omega t, \label{Damp_kernel}
\end{equation}
where  $\gamma(\omega)$ is the coupling strength
\begin{equation}\label{gamma_omega}
\gamma(\omega)=\sum_i \epsilon_i^2\,\delta(\omega-\omega_i),
\end{equation}
and by a noise kernel
\begin{equation}
 \nu(t) = \frac{1}{2}\average{\left\{\eta(t),\eta(0)\right\}} = \int_0^\infty \rmd\omega\;\frac{\gamma(\omega)}{\omega^2}\, E(\omega,T)\,\cos \omega t.\label{eq:noiseKer}
\end{equation}
In~\ref{app:equivalence} we show the equivalence between Ullersma's expression (\ref{H-solutions}) and Fleming's one (\ref{Lang_sol},\ref{Qi_t_formal}) for the solution of the QLE.

Having determined the time evolution of the Heisenberg momenta and positions as functions of the same operators at time $t=0$, all the moments of these quantities at time $t$ can now be evaluated as functions of the moments at $t=0$ and of the $A_{\mu\nu}$'s.

%--------------------------------------------------------------------
\subsection{Initial conditions}\label{ic}

General initial conditions were specified in equation (\ref{factor_ic}).
An equivalent description of the system can be obtained via the Wigner quasi-probability distribution (often simply called ``Wigner''), a function of the phase-space variables $(q,p)=(q_{0},p_{0},\ldots,q_{N},p_{N})$, defined in term of the total density matrix $\rho$ by
\begin{equation}\label{W_S}
\fl W(q,p)=\frac{1}{(\pi\hbar)^N}\int_{-\infty}^\infty \prod_{\mu}\rmd y_{\mu}\; \rme^{\rmi p_{\mu}y_{\mu}/\hbar} \bra{q_{0}-\frac{y_{0}}{2},\ldots,q_{N}-\frac{y_{N}}{2}}\rho\ket{q_{0}+\frac{y_{0}}{2},\ldots,q_{N}+\frac{y_{N}}{2}}.
\end{equation}
The reduced Wigner corresponding to the central oscillator and the bath reduced density matrix can be defined in a similar way and can be obtained from the total system Wigner by integrating out the appropriate degrees of freedom. 

Using a matrix formalism with vectors $\tilde{\bm{z}}^\mathsf{T}=(p,q)$, ${\bm{z}}^\mathsf{T}=(Q,P)$ and $\bm{k}^\mathsf{T}=(k_q,k_p)$, a generic single-particle Gaussian Wigner and its Fourier transform read \cite{Agarwal}:
\begin{eqnarray}\label{Gauss-Wig}
&&W(q,p) = \frac{1}{\sqrt{2\pi \Delta^2}}\exp{\{-\frac{(\tilde{\bm{z}}-\average{\tilde{\bm{z}}})^{\mathsf{T}}\bm{\sigma}(\tilde{\bm{z}}-\average{\tilde{\bm{z}}})}{2\Delta^2} \}}; \\
&&\label{Gauss-Wig-Fou}\widetilde{W}(\bm{k}) = \exp{\{-\frac{1}{2}\bm{k}^{\mathsf{T}} \bm{\sigma} \bm{k}-\rmi\bm{k}^{\mathsf{T}}\average{\bm{z}} \}}.
\end{eqnarray}
The first moments and the symmetric covariance matrix respectively read $\average{\tilde{\bm{z}}}^\mathsf{T}=(\average{P},\average{Q})$, $\average{\bm{z}}^\mathsf{T}=(\average{Q},\average{P})$ and
$\bm{\sigma}_{ij}=\average{\{\bm{z}_i,\bm{z}_j\}}/2-\average{\bm{z}_i}\average{\bm{z}_j}$ with $i,j=1,2$,
where we denote by $\{\ldots,\ldots\}$ the anticommutator and by $\average{\ldots}$ the average over a Gaussian density matrix $\rho$. 
We also define
\begin{eqnarray}\label{Sigma-Wigbis}
\Delta=\left(\sigma^2_{q}\sigma^2_{p}-C_{qp}^2\right)^{\frac{1}{2}}=(\det\bm{\sigma})^{\frac{1}{2}},
\end{eqnarray}
where we indicate $\sigma^2_{q}=\bm{\sigma}_{11}$, $\sigma^2_{p}=\bm{\sigma}_{22}$ and $C_{qp}=\bm{\sigma}_{12}$.

We shall only consider initial conditions such that the Wigner of the central oscillator has a Gaussian expression at time $t=0$. 
Then the Wigner is parametrized by its moments $\average{Q(0)}$, $\average{P(0)}$ $\average{Q^2(0)}$, $\average{P^2(0)}$ and $C_{qp}(0)=\average{\{Q(0)-\average{Q(0)},P(0)-\average{P(0)}\}}/2$. 
Since the initial density matrix of the bath is a product of exponentials of quadratic Hamiltonians, its corresponding Wigner is a product of Gaussian states which are parametrized, for $i=1,\ldots,N$, by the moments
\begin{eqnarray}\label{bath_ic}
&&\average{Q_i(0)} = \average{P_i(0)}=\average{\{Q_i(0),P_i(0)\}}=0, \\
&&\average{Q_i^2(0)}=\frac{E(\omega_i,T)}{\omega_i^2} \ \ , \ \ \average{P_i^2(0)}=E(\omega_i,T), \nonumber
\end{eqnarray}
where
\begin{equation}
E(\omega,T)=\frac{\hbar \omega}{2}\coth{\frac{\hbar\omega}{2 T}}.
\end{equation}
As a result the initial total Wigner is also Gaussian.

%--------------------------------------------------------------------
\subsection{Evolution} \label{ME}

At time $t>0$ the total density matrix operator will evolve in a unitary way (\ref{rho_tot_ev})
and the central oscillator and the bath will be correlated.
The corresponding total Wigner satisfies the Liouville-like evolution equation~\cite{Haake, Zurek}
\begin{equation}
\frac{\partial}{\partial t}W(q, p,t)=\Pb{H}{W}, 
\end{equation}
where $H$ (\ref{Hamilt}) is now considered as a function of the phase-space variables $(q,p)=(q_{0},p_{0},\ldots,q_{N},p_{N})$ and where $\Pb{\ldots}{\ldots}$ are the Poisson brackets. Again, by the linearity of the dependence of the solution for $(q,p)$ on the initial conditions, an initial Gaussian distribution remains Gaussian at later times. This means that the Wigner is a real Gaussian, positive definite at all times, and fully characterized by its first and second moments. This also applies to the bath and will be useful to evaluate its entropy as we are going to see in section \ref{bath_entropy}.  

It was shown in refs.~\cite{Haake,Fleming} that the reduced Wigner satisfies the following partial differential equation:
\begin{equation}\label{Wig_me}
\frac{\partial}{\partial t}W_\sys({\bm z},t) = \left[\nabla_{\bm{z}}^{\mathsf{T}}\cdot\bm{\mathcal{H}}(t)\cdot\bm{z}+\nabla_{\bm{z}}^{\mathsf{T}}\cdot{\bm D}(t)\cdot\nabla_{\bm{z}}\right]W_\sys(\bm{z},t),
\end{equation}
where the \emph{pseudo-Hamiltonian} ${\bm{\mathcal{H}}}(t)$ and \emph{diffusion} ${\bm D}(t)$  matrices reported in (\ref{ME-Matr})
 depend on the coupling strength respectively via the damping and noise kernels.  

The solution of equation (\ref{Wig_me}) can be found by a Fourier transformation, via the method of characteristics~\cite{Fleming}:
\begin{equation}\label{W_S_Fourier}
\widetilde{W}_\sys(\bm{k},t)=\widetilde{W}_\sys({\bm\Phi} ^{\mathsf{T}}(t)\bm{k},0)\,\rme^{-\frac{1}{2}\bm {k} ^{\mathsf{T}}\bm {\sigma}_T(t)\bm {k}},
\end{equation}
which appears as a product of a function depending on the Wigner at time zero $\widetilde{W}_\sys({\bm k},0)$, times a Gaussian one containing the thermal covariance. It clearly assumes Gaussian form in our hypotheses where the initial $\tilde{W}_S({\bm k},0)$ is Gaussian (\ref{Gauss-Wig-Fou}). The dynamics of the central oscillator is then fully described by the first and second moments of the position and momentum operators:
\begin{eqnarray}\label{mom_co_t}
&&\average{\bm{z}(t)} =\bm{\Phi}(t)\bm{z}_0;\\
&&\bm{\sigma}(t) =\bm{\Phi}(t)\bm{\sigma}_{0}\bm{\Phi}^{\mathsf{T}}(t)+\bm{\sigma}_T(t).
\end{eqnarray}
The general covariance matrix ${\bm \sigma}(t)$ corresponds to the covariance matrix in (\ref{Gauss-Wig}) if the averages $\average{\ldots}$ are evaluated with the total density operator at time $t$ (\ref{rho_tot_ev}). It appears as the sum of the contribution of the evolution of the initial conditions and of the thermal covariance:

\begin{equation}\hspace{-2cm}
\bm{\sigma}_T(t)=\int_0^\infty \rmd\omega\;\frac{\gamma(\omega)}{\omega^{2}}\,E(\omega,T) 
\left[
\begin{array}{cc}
\left|\int_0^t\rmd t'\;A(t')\rme^{i\omega t'}\right|^2 & \frac{1}{2}\frac{\rmd}{\rmd t}\left|\int_0^t\rmd t'\;A(t')\rme^{\rmi\omega t'}\right|^2\\
\frac{1}{2}\frac{\rmd}{\rmd t}\left|\int_0^tdt'A(t')\rme^{\rmi\omega t'}\right|^2 &  \left|\int_0^t\rmd t'\;\dot{A}(t')\rme^{\rmi\omega t'}\right|^2
\end{array}
\right]. \label{CM_sist}
\end{equation}

 The same expressions can be found by taking the average over initial conditions of the operators in the Heisenberg form (\ref{H-solutions}), expressing the $A_{i0}(t)$'s as functions of $A(t)$ (\ref{A_i0-A}) and by then using the coupling strength (\ref{gamma_omega})~\cite{Haake}. 
The elements of the correlation matrix are reported in more detail in \ref{num-cov-matr}.

At finite sizes one expects oscillatory behavior both for the dissipation and the diffusion coefficients. As in Ref.~\cite{Ullersma196627}, in the following we are going to assume an Ohmic form with a large cut-off for the couplings $\epsilon_i^2$ in (\ref{gamma_omega}). This choice enables us to obtain time-independent dissipation coefficients in the continuum frequency limit, while the diffusion ones only become time-independent in certain limits such as the high temperature limit.  In general, however, this would not be the case: by assuming for example a sub-Ohmic coupling with a slower decay for larger frequencies, one would have time-dependent and nonlocal dissipation and diffusion coefficients at all times, even in the high-temperature limit~\cite{Fleming}.

%%%%%%%%%%%%%%%%%%%%%%%%%%%%%%%%%%%%%%%%%%%%%%%%%%%%%%%%%%%%%%%%%%%%%%
\section{Thermodynamic limit} \label{therm_limit}

The thermodynamic limit of an infinite number of bath oscillators is obtained by substituting a continuous function $\gamma(\omega)$ to the discrete coupling strength (\ref{gamma_omega}). We choose the Drude-like Ullersma coupling strength~\cite{Haake, Ullersma196627}
\begin{equation}\label{U_strenght}
\gamma(\omega)=\frac{2}{\pi}\frac{\kappa\alpha^{2}\omega^2}{\alpha^2+\omega^2}.
\end{equation}
The parameter $\kappa$ tunes the strength of the coupling, while the cut-off $\alpha$, which is introduced in order to eliminate ultra-violet divergences, can be associated to the bath memory time. In fact the damping kernel (\ref{Damp_kernel}) with this coupling strength is given by 
\begin{equation}
K(t)=\kappa\alpha\, \rme^{-\alpha t},
\end{equation}
and thus decays over times of order $\alpha^{-1}$.

In this situation, the renormalized frequency $\Omega_{0}$ (\ref{eq:omega0}) of the central oscillator is simply given by
\begin{equation}
\Omega_{0}^{2}=\omega_0^2-\kappa\alpha.
\end{equation}

The general form of the propagator $A(t)$ following from the strength (\ref{U_strenght}) can be found in \cite{Haake}. 
It is characterized by three time scales: $\Omega$, $\Gamma$ and $\lambda$, deriving from the poles $\lambda$, $\Gamma\pm i\Omega$ of the Laplace transform of the propagator (\ref{Lap_A}). 

The propagator $A(t)$ describes a noisy damped oscillator, where $\Omega$ is the characteristic frequency and $\lambda$ and $\Gamma$ characterize the damping rates. 
When the time scale $1/\lambda$ is much shorter than $1/\Gamma$ and $1/\Omega$, the damping kernel $K(t)$ (\ref{Damp_kernel}) becomes delta-like and the QLE (\ref{Lang-Matr}) becomes local in time. This is obtained by taking the large cut-off limit, defined by 
\begin{equation}\label{high-cut-off}
 \alpha\gg\kappa,\omega_0.
\end{equation}
In this limit the propagator $A(t)$ assumes the form
\begin{equation}\label{A_local}
 A_{\loc}(t)=\frac{1}{\Omega}\sin(\Omega t)\rme^{-\Gamma t},
\end{equation}
which is typical of a damped Ornstein-Uhlenbeck process. The equations that determine $\Gamma$, $\Omega$ and $\lambda$ in the general case
and in the limit (\ref{high-cut-off}) are reported in~\ref{Timescales}, which also describes the transition between the under-damped and over-damped dynamics.

The quantum and time-dependence features of our process are then contained only in the noise kernel (\ref{eq:noiseKer}),
\begin{eqnarray}\label{noise_kern}
\nu(t)=-\kappa\alpha^2\left(\frac{1}{2}\cot\frac{\beta\alpha}{2}\rme^{-\alpha t}+\frac{1}{\pi}\sum_{\ell=1}^\infty \frac{ \ell}{(\alpha\tau_{\beta})^2-\ell^2}\rme^{-\ell t/\tau_{\beta}}\right),
\end{eqnarray}
where we have defined
\begin{equation}\label{eq:taubeta}
\tau_{\beta}=\frac{\hbar\beta}{2\pi}.
\end{equation}

We can thus define the following limits:
\begin{itemize}
\item[] \textbf{The low-temperature limit:}
\begin{equation}\label{eq:LT-limit}
\alpha\gg  1/\tau_{\beta} ;
\end{equation}
\item[] \textbf{The high-temperature classical limit:}
\begin{equation}\label{CL-limit}
1/\tau_{\beta} \gg \alpha ;
\end{equation}
\item[] \textbf{The weak-coupling limit:}
\begin{equation}\label{eq:WC-limit}
\Gamma\ll\Omega,1/\tau_\beta.
\end{equation}
\end{itemize}

The noise kernel determines the thermal covariance matrix $\bm{\sigma}_{T}$ (\ref{thermal_M}) and, via (\ref{eq:diffusion}), the diffusion coefficients of the ME. Thus the quantum and time-dependence features will show up in these quantities.
One can evaluate the covariance matrix by using  the local propagator (\ref{A_local}) inside the general expression (\ref{CM_sist}), up to terms of~$\Order{1/\alpha}$.
Using this propagator instead of the general one (see~\cite[eq.(7.10)]{Haake})
does not affect either the covariance thermal matrix or the diffusion coefficients in the large cut-off limit, even for times $t<1/\alpha$, since only correction of~$\Order{1/\alpha}$ arise~\cite{Fleming}. Since the coupling strength (\ref{U_strenght}) is an even meromorphic function, the integrals appearing in (\ref{CM_sist}) can be evaluated by a contour integration in the complex plane.

Complete expression for the thermal correlation matrix in the large cut-off limit, which are exploited in the following for the calculation of the entropies, were derived in~\cite{Haake} and are reported in~\ref{num-cov-matr}.  Their quantum features are due to the presence of the function $E(\omega,T)$, whose poles at $\omega=\rmi k\,2\pi /\tau_{\beta}$ with $k$ any integer, give rise to the thermal transients, i.e., to terms which vanish on a time scale of order $\tau_{\beta}$. These terms are also responsible for the time-dependence of the diffusion coefficients of the master equation~\cite{Haake}. In the high-temperature classical limit, where $E(\omega,T)$ approaches $T$, all the thermal transients vanish and the expressions of the covariance matrix simplify.

One can deduce from equations (\ref{M_prop}), (\ref{W_S_Fourier}) and (\ref{A_local}) that, since $\lim_{t\to\infty}A_{\loc}(t)=0$, in the thermodynamic limit the system eventually loses all information on its initial conditions, and its distribution assumes the characteristic Gaussian form corresponding to the late-time thermal covariance matrix, as described in the next subsection.

An important feature of this model is the presence of initial slips in the momentum average, in the non-thermal part of the averaged square momentum and of the correlation between $Q$ and $P$. Using
the local propagator (\ref{A_local}) from $t=0$, implies neglecting an initial evolution of the system during a short time of order $1/\alpha$, in which the central oscillator is subjected to an initial kick~\cite{Fleming,Haake}.
One can easily observe, indeed, that
\begin{equation}\label{ddA_slip}
 \ddot{A}(0)=0\neq\ddot{A}_{\loc}(t{=}0^{+})=-2\Gamma.
\end{equation}
Our description will thus only be valid for $t\gg1/\alpha$. The effect of initial slips both on the moments and on the definitions of entropy production is discussed in \ref{sec:slips}.  

%---------------------------------------------------------------
\subsection*{Late-time covariance matrix} \label{Late-cov-mat}

In the thermodynamical limit it is possible to evaluate the long-time behavior of the diffusion coefficients and of the covariance thermal matrix. They are related by
\begin{eqnarray}\label{Dqp}
 &&D_{qp}(\infty) =\average{P^2(\infty)}-\Omega_0^2\average{Q^2(\infty)};\\
&&D_{pp}(\infty) =2\Gamma\average{P^2(\infty)}.
\end{eqnarray}
Thus the anomalous diffusion coefficients survive, since the right-hand side of the first equation of (\ref{Dqp}) does not vanish. This implies that equipartition does not hold in the general quantum case.

Interestingly, as observed in~\cite{Haake}, one obtains
\begin{eqnarray}\label{M2eq}
\average{Q^2(\infty)} =\average{Q^2}_{\eq} \ \ , \ \ \average{P^2(\infty)} =\average{P^2}_{\eq},
\end{eqnarray}
namely that the stationary form of the central oscillator density matrix at $t=\infty$ equals the traced canonical equilibrium one of the total system:
\begin{eqnarray}\label{rho_st_eq1}
\rho_\sys(\infty) =\rho^{\st}_{\sys}= \tr_\B \rho^{\eq} \ \ , \ \ 
\rho^{\eq} \equiv \frac{\rme^{-\beta H}}{Z} \ \ , \ \ 
Z=\tr \rme^{-\beta H}.
\end{eqnarray}
This does not mean of course that the total system equilibrates: $\rho(\infty) \neq \rho^{\eq}$~\cite{Hilt}.
Furthermore, it has been shown in~\cite{Grabert} that 
\begin{eqnarray}\label{rho_st_eq}
\rho_\sys(\infty) = \rho_\sys^{\eq} \equiv \frac{\rme^{-\beta H_\sys^{\eq}}}{Z_\sys^{\eq}} \ \ , \ \
Z_\sys^{\eq}=\tr_\sys \rme^{-\beta H_\sys^{\eq}},
\end{eqnarray}
where the equilibrium effective Hamiltonian $H_\sys^{\eq}$ is given by
\begin{equation}\label{H_S_eq}
H_\sys^{\eq}=\frac{1}{2M_{\eff}}P^2+\frac{1}{2}M_{\eff}\omega_{\eff}^2Q^2.
\end{equation}
The effective frequency $\omega_{\eff}$ and mass $M_{\eff}$ are respectively given by
\begin{eqnarray}
\omega_{\eff} =\frac{2}{\beta\hbar}\coth^{-1}\left(\frac{2}{\hbar}\sqrt{\average{Q^2}_{\eq}\average{P^2}_{\eq}}\right) \ \ , \ \
M_{\eff} =\frac{1}{\omega_{\eff}}\sqrt{\frac{\average{P^2}_{\eq}}{\average{Q^2}_{\eq}}}.
\end{eqnarray}
Expressions for $\average{Q^2}_{\eq}$ and $\average{P^2}_{\eq}$ can be found in \cite{Grabert} and~\cite{Haake}, and are reported in (\ref{Q2-eq},\ref{P2-eq}).

It is worth noticing that the traced canonical equilibrium density matrix (\ref{rho_st_eq}) can be equivalently written in the form \cite{HanggiIngoldTalkner08NJP, HanggiIngoldActaPol06}:
\begin{equation} \label{RhoCanEff}
 \rho_\sys^{\eq}=\frac{\rme^{-\beta H_\sys^{\MF}}}{Z_\sys^{\MF}} \ \ , \ \ 
H_\sys^{\MF}=-\frac{1}{\beta}\ln\frac{\tr e^{-\beta H}}{Z_\B} \ \ , \ \
  Z_\sys^{\MF}=\frac{Z}{Z_\B}\ \ ,
\end{equation}
where an Hamiltonian of mean force $H_\sys^{\MF}$ has been introduced, which differs from the effective $H_\sys^{\eq}$ by the additive constant $-\beta(\ln Z_\sys^{\MF}-\ln Z_\sys^{\eq})$. Both partition functions have a well-known analytical expression \cite{Grabert}. For the practical purpose of evaluating the Breuer entropy flow (\ref{DS_e_eq}) we will use the effective Hamiltonian. However we emphasize that the use of the mean force Hamiltonian leads to exactly the same entropy production (since only the density matrix is involved) and heat flow (since only differences in energies are considered). 

The identities (\ref{M2eq},\ref{rho_st_eq1},\ref{rho_st_eq}) do not generally hold in open quantum systems. They are however an important feature of our bilinear model and hold independently of the choice of the continuous limit strength.

%%%%%%%%%%%%%%%%%%%%%%%%%%%%%%%%%%%%%%%%%%%%%%%%%%%%%%%%%%%%%%%%%%%%%%
\section{Explicit forms of the entropy production}\label{Results}
%%%%%%%%%%%%%%%%%%%%%%%%%%%%%%%%%%%%%%%%%%%%%%%%%%%%%%%%%%%%%%%%%%%%%%

We report here the explicit forms of the entropy production, according to the $\P$
(\ref{DS_i_poi}), the ELB (\ref{DS_i_esp}) and the `Breuer' expression obtained in (\ref{DS_i_eq}). To evaluate them one needs to know the expressions of the entropy and the entropy flow.

Since the central oscillator density matrix is Gaussian at each time $t$, its von~Neumann entropy entropy is given by~\cite{Agarwal} 
\begin{eqnarray}\label{Sist_entr_gauss}
S(t)&=& -\tr_\sys\rho_\sys(t)\ln\rho_\sys(t)\nonumber\\
&=&\left(\Delta(t)+\frac{1}{2}\right)\ln\left(\Delta(t)+\frac{1}{2}\right)-\left(\Delta(t)-\frac{1}{2}\right)\ln\left(\Delta(t)-\frac{1}{2}\right),\end{eqnarray}
where we have defined
\begin{equation}
 \Delta(t) =\hbar^{-1}\left(\sigma_q^2(t)\sigma_p^2(t)-C_{qp}^2(t)\right)^{{1}/{2}},
\end{equation}
which is a function of the correlation matrix at time $t$. One notices that $S(t)$ is well defined if the uncertainty principle is satisfied.

The `Poised' entropy production can be written as
\begin{eqnarray}
&&\Delta_{\rm i} S^{\P}=\Delta S-\Delta_{\rm e} S^{\P};\\
&&\Delta_{\rm e} S^{\P}=\tr\left(\rho_{S}(0)-\rho_{S}(t)\right)\ln\rho_{S}^{*}(t).
\end{eqnarray}
The `Poised' density matrix $\rho^{*}_{S}(t)$ is Gaussian with vanishing means of $Q$ and $P$ (as shown in~\ref{app:breuer}) and is given by
\begin{equation}\label{ln_poised}
\ln\rho_{S}^{*}(t)=-\frac{1}{2}\ln\left(\Delta^{*2}(t)-\frac{1}{4}\right)-\frac{\Lambda^{*}(t)}{2\hbar^2\Delta^{*}(t)}\ln\frac{\Delta^{*}(t)+\frac{1}{2}}{\Delta^{*}(t)-\frac{1}{2}} ,
\end{equation}
where we have defined
\begin{eqnarray}
&&\Delta^{*}(t)=\hbar^{-1}\left({\sigma^{*}_q}^2(t){\sigma^{*}_p}^2(t)-{C^{*}_{qp}}^2(t)\right)^{{1}/{2}};\\
&&\Lambda^{*}(t)={\sigma^{*}_{p}}^{2}(t)Q^{2}+{\sigma^{*}_{q}}^{2}(t)P^{2}-C^{*}_{qp}(t)\{Q,P\}.
\end{eqnarray}
The variances and correlation ${\sigma^{*}_{q,p}}^{2}(t)$ and $C^{*}_{qp}(t)$ are given in~\ref{app:breuer}. Then one obtains
\begin{eqnarray}
\Delta_{\rm e} S^{\P}=\frac{\average{\Lambda^{*}(t)}_{t}-\average{\Lambda^{*}(t)}_{0}}{2\hbar^2\Delta^{*}(t)}\ln\frac{\Delta^{*}(t)+\frac{1}{2}}{\Delta^{*}(t)-\frac{1}{2}},
\end{eqnarray}
where we have defined, for any operator $O$ acting on the Hilbert space of S, $\average{O}_{t}=\tr \rho_{\sys}(t) O$.

As for the Breuer entropy flow, $\Delta_{\rm e} S^{\mathrm{\Br}}$ is straightforwardly given by the change in the effective energy $H^{\eq}_{\sys}$ (\ref{H_S_eq}). Then one only needs to know position and momentum second moments at time $t$, which in the finite case are obtained from the first moments and from the correlation matrix which appear in~(\ref{mom})-(\ref{sigma-therm-discr-p2}), while in the continuum case one exploits the general expressions (\ref{mom_co_t})-(\ref{CM_sist}) with the Ullersma coupling strength (\ref{U_strenght}) (see (\ref{sigma-therm}) and (\ref{sigma-therm-P2})). The entropy flow $\Delta_{\rm e} S$ is instead proportional to the change in the bath energy (\ref{DS_e_esp}). To evaluate it, one needs rather to evaluate the average of the interaction energy term $\average{H_\mathrm{I}}$. 
By using the Ullersma strength (\ref{U_strenght}) in the large cut-off limit, it turns out that for $t\gg 1/\alpha$
\begin{equation}\label{H_I_t}
 \average{H_I(t)}=D_{qp}(t)-\kappa\alpha\average{Q^2(t)},
\end{equation}
where $D_{qp}(t)$ is the anomalous diffusion coefficient (\ref{eq:diffusion}).
This evaluation is reported in \ref{A-Int-term}.
Thus, by comparing definitions (\ref{DS_e_esp}) and (\ref{DS_e_eq}), we obtain the 
difference between the two entropy flows is given by
\begin{equation}\label{Diff_DSe}
 \Delta S_e-\Delta_{\rm e} S^{\Br}=\beta\left[\Delta\average{H_\sys}-\Delta\average{H_\sys^{\eq}}-\kappa\alpha\average{Q^2(t)}+D_{qp}(t)\right].
\end{equation}
The difference of entropy production is the same with opposite sign.

As already mentioned, the expressions we use in the continuum limit for the three definitions of entropy only apply for $t\gg1/\alpha$, after the initial slip has taken place. Their contribution to entropy, which is reported in~\ref{sec:slips}, implies that the entropy flows and productions often do not start from $0$, as one can observe in the following figures.

%------------------------------------------------------------
\subsection{The Markovian case}\label{Mark_Kramers}

Generally, in the limit of short-lived thermal transients, namely $\Gamma\tau_{\beta}\ll 1$, the generator of the dynamics $\mathcal{L}$ can be considered time-independent, since the diffusion coefficients (\ref{eq:diffusion}) are close to their $t=\infty$ limit~(\ref{Dqp}). As we have seen in section \ref{BreuerSect}, in this case the entropy production definition (\ref{DS_i_eq}) and its time derivative turn out to be consistently positive, as the system equilibrium density matrix $\rho_\sys^{\eq}$ does not depend on time, and the entropy flow is given by the average variation of the effective system Hamiltonian $H_\sys^{\eq}$, as already observed in equation~(\ref{DS_e_eq}). This holds both in the high-temperature classical and weak-coupling limits \cite{Haake}.

\subsubsection{Classical limit.}\label{Res-CL}

In the high-temperature classical limit (\ref{CL-limit}) all the quantum features of the system disappear and the anomalous diffusion coefficient vanishes, thus recovering equipartition (\ref{Dqp}), since $\average{P^2}_{\eq}=T$, and $\average{Q^2}_{\eq}=T/\Omega_0^2$. In particular the equation satisfied by the Wigner has exactly the form of the Kramers equation for an oscillator in contact with a bath at temperature $T$ \cite{Kramers40, Risken}. In this limit, since we have $\Delta(t) \gg 1$, $\forall t$, the system entropy (\ref{Sist_entr_gauss}) assumes its classical form for a Gaussian distribution
\begin{equation}
 S(t) \simeq 1+\ln \Delta(t),
\end{equation}

\begin{figure}[!h]\centering
\includegraphics[width=\textwidth]{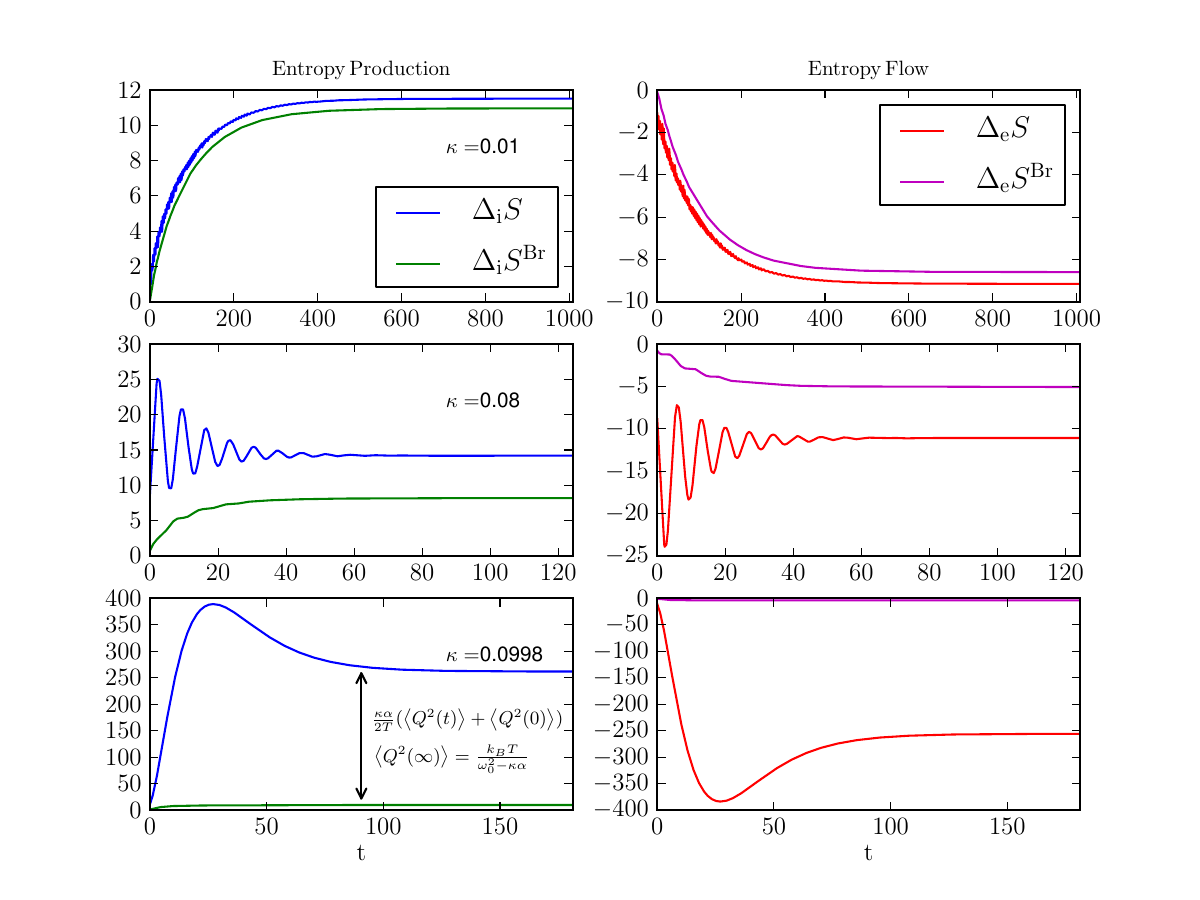}
\caption{The different expressions for the entropy production (left) and the entropy flow (right), for different couplings $\kappa$ (top and center underdamped, bottom overdamped) in the classical regime. The parameters are: temperature $T=1000$, $\alpha=10$, $\omega_0=1$, $\sigma^2_q(0)=100$, $\sigma^2_p(0)=100$, $C_{qp}(0)=10$, $\average{Q(0)}=100$, $\average{P(0)}=100$.}
 \label{fig:CL_therm_split} 
\end{figure}

Moreover the effective equilibrium energy is given by
\begin{equation}\label{H_S_eq_CL}
 H_\sys^{\eq}\simeq\frac{1}{2}(\Omega_0^2Q^2+P^2).
\end{equation}
Thus the definition (\ref{DS_e_eq}) of the entropy flow reduces to that of stochastic thermodynamics, which is defined as the average variation of the effective energy of the system, namely the classical one with the renormalized frequency $\Omega_0$ in place of $\omega_0$. This means that the corresponding definition of the entropy production coincides in this limit with the one introduced in the theory of stochastic thermodynamics for the Kramers equation~\cite{Imparato}. In the overdamped limit $\Gamma\gg\Omega_0$ the momentum equilibrates much faster than position and can thus be traced out. The entropy production assumes in this case the form proposed in the theory of stochastic thermodynamics for the overdamped Fokker-Planck equation~\cite{Seifert}. As long as momentum has not yet fully equilibrated, the latter expression constitutes a lower bound to the former one since it results from a coarse graining procedure (see, e.g., \cite{VdBParrondoPRE08}). 
Let us also note that by taking the weak coupling limit $\kappa\rightarrow 0$ one gets in (\ref{H_S_eq_CL}) the bare frequency $\omega_0$, and that then the entropy flow becomes exactly equal to the change in the central oscillator energy, divided by the temperature of the bath. 

We show in Figure~\ref{fig:CL_therm_split} the high-temperature limit (\ref{CL-limit}) of the difference between the different definitions of the entropy production with different coupling strengths. In the classical limit the anomalous diffusion term in (\ref{H_I_t}) vanishes and the normal diffusion coefficient is time independent. This means that the Poised and Breuer expressions for entropy become equal: $\Delta_{\rm i} S^{\Br}=\Delta_{\rm i} S^{\P}$.
Considering also the expression assumed by $H_\sys^{\eq}$ (\ref{H_S_eq_CL}), the expression~(\ref{Diff_DSe}) for the difference of flows simplifies to
\begin{equation}
\Delta_{\rm e} S-\Delta_{\rm e} S^{\Br}=-\frac{\beta}{2}\kappa\alpha\left(\average{Q^2(t)}+\average{Q^2(0)}\right).
\end{equation}
The same difference with opposite sign holds for the entropy production.
In the classical limit the thermal part of $\average{Q^2(t)}$ is proportional
to $1/\beta\Omega_0^2$. This means that the difference between the two expressions for the entropy production \textit{diverges}, since for large $\kappa$ one has
\begin{equation}\label{diver_limit}
 \Omega_{0}^{2}=\omega_{0}^{2}-\kappa\alpha\rightarrow 0.
\end{equation}

This appears clearly in the figure, where the different expressions for the entropy production $\Delta_{\rm i} S$ and for the entropy flow $\Delta_{\rm e} S$ are shown for different coupling strengths $\kappa$, both in the underdamped and the overdamped regime. 

The difference between the definitions is due to the fact that the expression $\Delta_{\rm i} S$ and the corresponding expression $\Delta_{\rm e} S$ of the entropy flow both diverge in the limit (\ref{diver_limit}) as $1/\Omega_0^2$,
\begin{equation}\label{DSe_infty_CL}
 \Delta_{\rm e} S(\infty)=1-\frac{1}{2}\frac{\kappa\alpha}{\Omega_0^2}-\frac{\beta}{2}\left(\omega_0^2\average{Q^2(0)}+\average{P^2(0)}\right).
\end{equation}
However, the expression $\Delta_{\rm i} S^{\Br}$ diverges only logarithmically like the von~Neumann entropy:
\begin{equation}\label{DS_infty_CL}
 \Delta S(\infty)\simeq|\ln\beta\Omega_0|-S(0)
\end{equation}
In fact the expression $\Delta_{\rm e} S^{\Br}$ does not diverge, since in the 
effective Hamiltonian (\ref{H_S_eq_CL}) only the renormalized frequency appears: $M_{\eff}\omega_{\eff}^2\rightarrow\Omega_0^2$.

One notices in~Figure~\ref{fig:CL_therm_split} that both expressions of the entropy productions are positive, but that the ELB one, $\Delta_{\rm i} S$, exhibits damped oscillations yielding a nonpositive time derivative. This can be directly seen from the fact that the time derivative
of $\Delta_{\rm i} S^{\Br}$ is positive, due to the fact that the process is time independent (cf.\ sec.~\ref{Mark_Kramers}), and that the ELB one differs from it by a constant plus a term proportional to $\average{Q^2(t)}$, which is characterized by damped oscillations.

We remark here that usually in literature the total Hamiltonian is renormalized by a self interaction term, such that no positivity condition similar to equation~(\ref{eq:omega0}) has to be satisfied.
 In this case there would not be any divergence of $\average{Q^2(t)}$, which would be proportional to $T/\omega_0^2$, but the difference between the two definitions of the
entropy production can be made arbitrarily large by taking $\kappa\rightarrow\infty$ \cite{Nieuwenhuizen}.

\begin{figure}[!h]
\centering
\includegraphics[width=\textwidth]{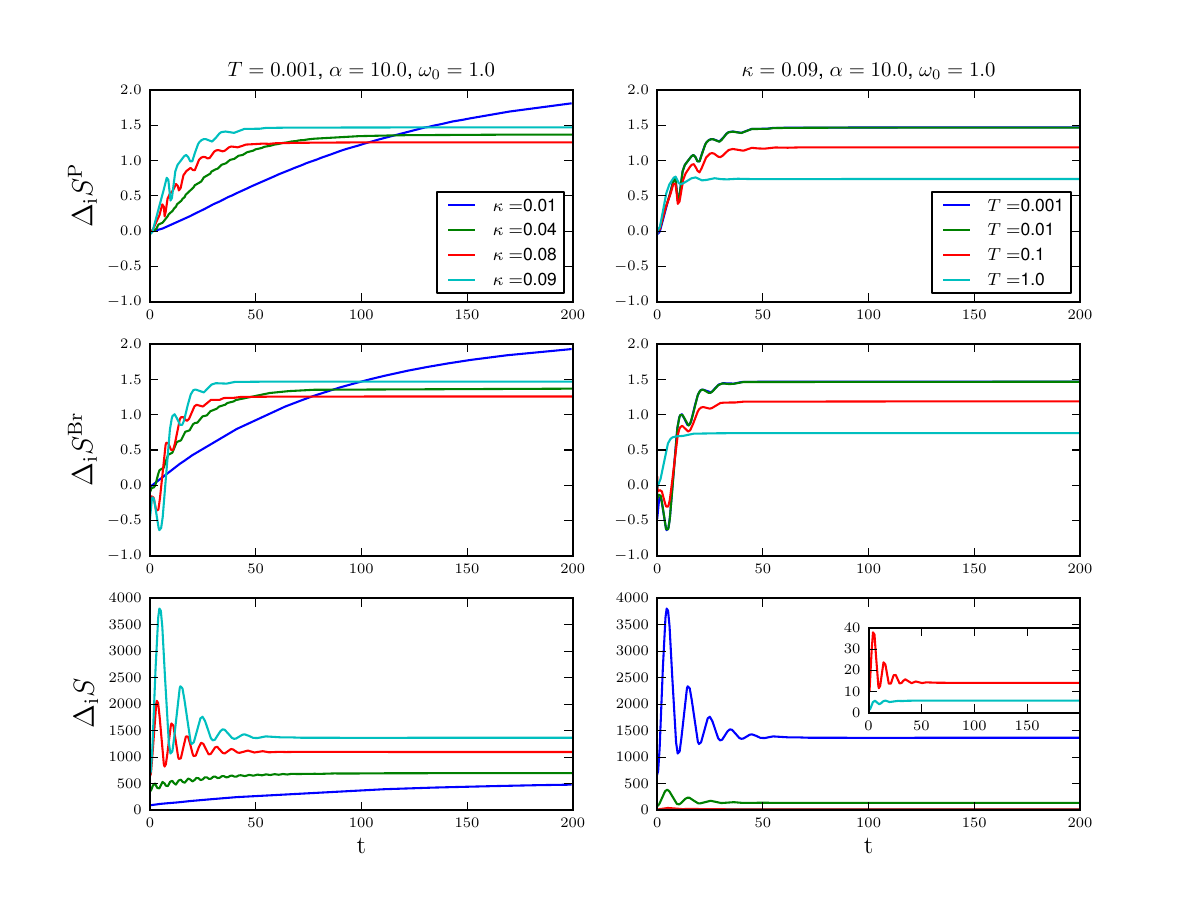}
\caption{Entropy productions $\Delta_{\rm i} S^{\P}$, $\Delta_{\rm i} S^{\Br}$ and $\Delta_{\rm i} S$  at different values of the coupling $\kappa$: $0.01$ (blue), $0.04$ (green), $0.08$ (red), $0.09$ (cyan) with temperature $T=0.001$ (left), and different temperatures $T$: $0.001$ (blue), $0.01$ (green), $0.1$ (red), $1.0$ (cyan), with $\kappa=0.09$ (right). Initial condition are fixed: $\sigma^2_q(0)=1.0$, $\sigma^2_p(0)=1.0$, $C_{qp}(0)=0$, $\average{Q(0)}=0$,
$\average{P(0)}=0$.}\label{fig:Quan_Br_Esp} 
\end{figure}

\subsubsection{Weak-coupling limit}\label{WC-lim}

Another case in which the entropy flow is the equal to the one defined in stochastic thermodynamics is the weak-coupling limit in the general quantum setting $\Gamma \ll \Omega,\tau_\beta^{-1}$. Some care is needed, since the anomalous diffusion coefficient $D_{qp}(t)$ does not vanish at long times to first order in the coupling $\Gamma$, just as the normal diffusion coefficient $D_{pp}(t)$:
\begin{eqnarray}
 &&D_{qp}(\infty) =\frac{2}{\pi}\hbar\Gamma\Re\left[\psi(1+\lambda\tau_{\beta})-\psi(1+i\Omega\tau_{\beta})\right]+\Order{\Gamma^2};\\
 &&D_{pp}(\infty) =2\Gamma E(\Omega,T)+\Order{\Gamma^2},
\end{eqnarray}
where $\psi(z)$ is the digamma function.
Anyway their contribution to $\average{Q^2}_{\eq}$ is different as $D_{pp}(t)$ contributes to order one, while $D_{qp}(t)$ to order $\Gamma$ as seen by inverting (\ref{Dqp}). One gets then equipartition to first order in $\Gamma$:
\begin{eqnarray}
\average{Q^2}_{\eq} =E(\Omega,T)/\Omega^2+\Order{\Gamma} \ \ , \ \
\average{P^2}_{\eq} =2\Gamma E(\Omega,T)+\Order{\Gamma},
\end{eqnarray}
where $\Omega$ can be approximated by $\Omega_0$ to first order in $\Gamma$.
This correspond to an equilibrium density matrix $\rho_\sys^{\eq}$ (\ref{rho_st_eq}) corresponding to the equilibrium Hamiltonian
\begin{equation}
H_\sys^{\eq}=\frac{1}{2}(\omega_0Q^2+P^2)
\end{equation}
which is the same as the central oscillator one (\ref{Hamilt}).

%--------------------------------------------------------------
\subsection{Low-temperature limit}

In the low-temperature limit (\ref{eq:LT-limit}) one expects that the Breuer entropy production expression (\ref{DS_i_eq}), as well as its time derivative, can become negative. The Poised and ELB expressions (\ref{DS_i_esp}) remain instead positive, while their time derivative can be negative. As we observe in Figure~\ref{fig:Quan_Br_Esp}, for sufficiently low temperature and strong couplings, the expression $\Delta_{\rm i} S^{\Br}$ becomes negative, exhibiting an oscillatory behavior. At higher temperatures or weaker couplings the amplitude of the oscillations becomes smaller. Thus in these limits one obtains a positive definite entropy production, as well as a positive time derivative.
We observe that the Poised and Breuer entropy production have the same asymptotic value, as expected, since $\rho_{\sys}^{*}(\infty)=\rho_{\sys}^{eq}$.

One notices that also in the low-temperature limit~(\ref{eq:LT-limit}) the ELB expression can be orders of magnitude larger than the other two, due to the coupling term $\kappa\alpha\average{Q^2(t)}$ which appears in the entropy flow.
This difference can be much larger respect to the classical case, due to the presence of the quantum terms contained in $\average{Q^2(t)}$, which actually become more relevant than the classical one.

%%%%%%%%%%%%%%%%%%%%%%%%%%%%%%%%%%%%%%%%%%%%%%%%%%%%%%%%%%%%%%%%%%%%%%
\section{Poincaré recurrences}\label{Poi_rec}
%%%%%%%%%%%%%%%%%%%%%%%%%%%%%%%%%%%%%%%%%%%%%%%%%%%%%%%%%%%%%%%%%%%%%%

When the number $N$ of bath oscillators is finite, the dynamics is
characterized by a recurrent behavior, with a period identified by
the Poincaré recurrence time $t_\mathrm{P}\sim 2\pi/\min(z_{\nu+1}-z_{\nu})$ \cite{Ullersma196627},
where the $z_{\nu}$'s are the normal frequencies. We can interpret this recurrence as an almost periodic return to the initial decoupled state. Interestingly, while $\Delta_{\rm i} S$ remains positive by definition,  one might have a negative $\Delta_{\rm i} S^{\Br}$, even in the classical case. When the size of the bath becomes larger, the recurrence time grows, and one expects that the entropy approaches its typical irreversible behavior, eventually relaxing to the equilibrium asymptotic value.

In the present section we study this behavior in the two specific cases of uniform and Lorentzian frequency sampling, always assuming that the coupling strength converges to the Ullersma expression (\ref{U_strenght}). Indeed, the density of states $\sum_i\delta(\omega-\omega_i)$ inside the coupling strength can be arbitrarily chosen. We evaluate the thermal covariance matrix components $\sigma_{q,T}^2(t)$, $\sigma_{p,T}^2(t)$, and the equilibrium symmetrized autocorrelation function $C(t)$, defined by
\begin{equation}
C(t)=\frac{1}{2}\average{\{Q(t),Q(0)\}}_{\eq}.
\end{equation}
We can also consider the Fourier transform of the correlation function $C(t)$. Indeed, in the classical limit, the finite-size correlation function has the expression
\begin{equation}
 C_N(t)= T\sum_{\nu=0}^N\frac{X_{0\nu}^2}{z_\nu^2}\cos(z_\nu t).
\end{equation}
We can thus represent the Fourier transform $\tilde{C}_{N}(\omega)$ of $C(t)$ by setting it equal to $TX_{0\nu}^2/(z_\nu^2\,\Delta_{\nu})$, where $\Delta_{\nu}=z_{\nu}-z_{\nu-1}$,  and considering it as a function of $\omega=z_{\nu}$. This quantity should approach, as $N\to\infty$, the Fourier transform of $C(t)$, which is given by
\begin{equation}
 \tilde{C}(\omega)= T\frac{\kappa\alpha^2/(2\pi)}{(\omega^2-\omega_0^2)^2(\alpha^2+\omega^2)+\kappa^2\alpha^4+2\kappa\alpha^3(\omega^2-\omega_0^2)}.
\end{equation}

We will see that the convergence to the large-size irreversible behavior is much slower for the uniform than for the Lorentzian sampling, and that, in the former case, the dynamics seems to remain characterized by underdamped oscillations even at large values of $N$.

%--------------------------------------------- 
\subsection{Sampling}
\label{sec:sampling}
%--------------------------------------------- 
% ..............................................................................
\subsubsection{Uniform.}
\label{subsubsec:Uniform}
% ..............................................................................
The uniform sampling is obtained by considering $N$ frequencies $\omega_{\ell}$ ($\ell=1,2,\ldots,N$) spaced by a constant $\Delta$. The maximal frequency $N\,\Delta$ will be denoted by $\omega_{\C}$. The corresponding couplings are given by
\begin{equation}
 \epsilon_\ell=\sqrt{\Delta\frac{2}{\pi}\frac{\kappa\alpha^2\omega_\ell^2}{\alpha^2+\omega_\ell^2}}.
\end{equation}
Then the continuous-limit Ullersma strength is obtained for $N\rightarrow\infty$,
 $\omega_c\rightarrow\infty$ and $\Delta=\omega_{\C}/N\rightarrow 0$.
In this case the Poincaré recurrence time is given by $t_P\simeq 2\pi/\Delta$.

% ..............................................................................
\subsubsection{Lorentzian.}
\label{subsubsec:Lorentzian}
% ..............................................................................
In order to obtain a faster convergence with longer Poincaré recurrence times, and a better agreement with the continuum curve both in the under-damping and in the over-damping cases, one can adopt a Lorentzian sampling of frequencies. Positive frequencies distributed with a Lorentzian density centered at $\omega=0$, with width $a_0$ are defined as
\begin{eqnarray}
 \omega_\ell =a_0\tan\left[\frac{\ell}{N+1}\frac{\pi}{2}\right],
\end{eqnarray}
with $\ell=1\ldots N$ and with the corresponding couplings
\begin{equation}
 \epsilon_\ell=\sqrt{\Delta_\ell\frac{2}{\pi}\frac{\kappa\alpha^2\omega_{\ell}^{2}}{\alpha^2+\omega_\ell^2}},
\end{equation}
where $\Delta_\ell=\omega_{\ell}-\omega_{\ell-1}$, $\ell=2,\ldots,N$ and $\Delta_{1}=\omega_{1}$.
This sampling enables a high density of frequencies in the area around $\omega=0$, then determining a long recurrence time. One can adjust the value of $\Delta_{N}$ in such a way as to have, for all values of $N$,
\begin{equation}\label{OM1_N_eq_kalph}
\sum_\ell\frac{\epsilon_\ell^2}{\omega_\ell^2}=\kappa\alpha.
\end{equation}
We shall refer to this case as the \textit{adjusted Lorentzian} sampling.
% ------------------------------------------------------------------------------
\subsection{Results}
\label{subsec:Results}
\begin{figure}[htp]
 \centering
 \includegraphics[width=\textwidth]{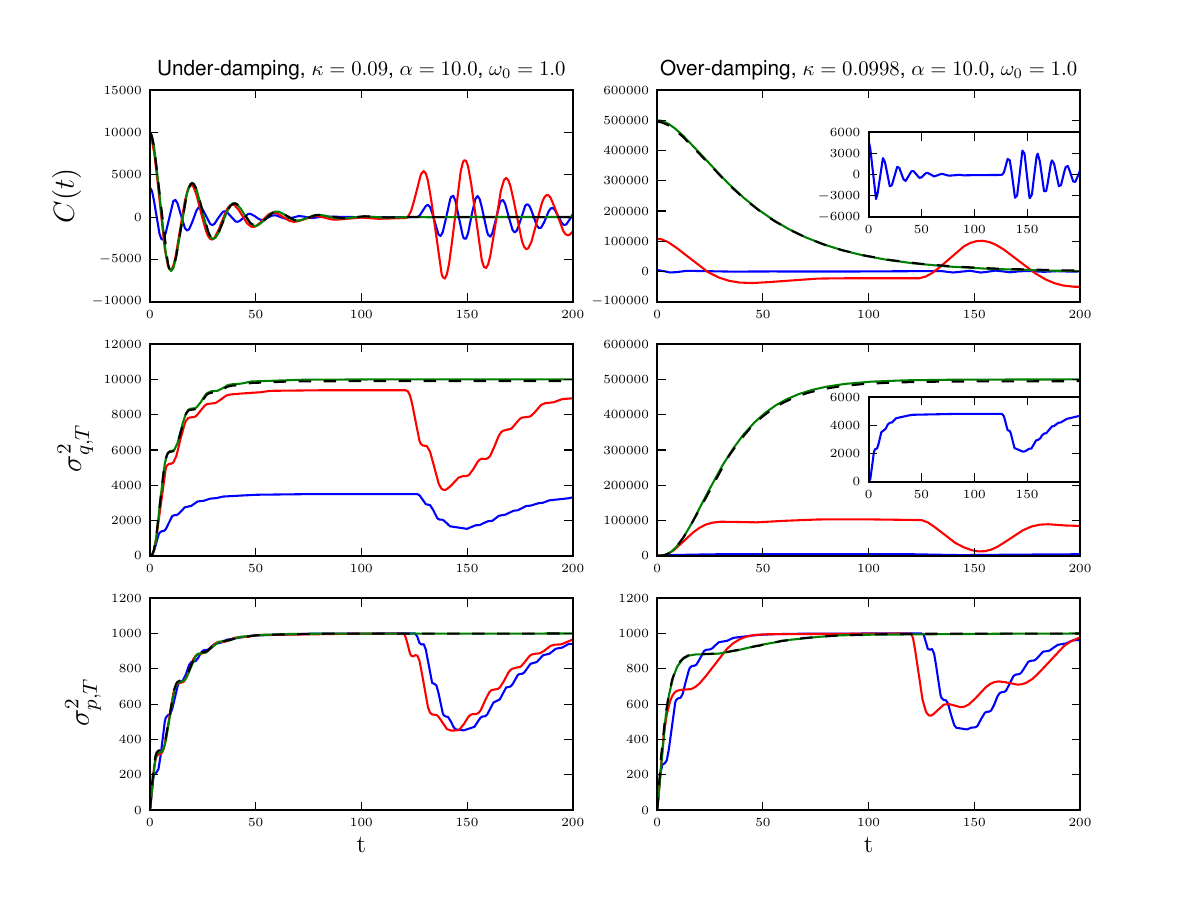}
 \caption{Plot of $C(t)$, $\sigma^{2}_{q,T}(t)$ and $\sigma^{2}_{p,T}(t)$ in the continuous limit (dashed), compared with the corresponding curves obtained for $N=600$ with a uniform frequency distribution with $\omega_\C=30.0$ (blue), a Lorentzian
distribution of frequencies with $a_0=20.0$, (red) and an adjusted Lorentzian distribution with $a_0=0.1$ (green). They are obtained both for an under-damping set of parameters (left column) and an over-damping one (right column) in the classical case $T=1000$. Insets are magnifications of the finite-size curves with uniform sampling.}\label{A_C_Q2_P2}
\end{figure}

% ------------------------------------------------------------------------------  
In Figure \ref{A_C_Q2_P2} we report the correlation function and the thermal part of second moments of the central oscillator in the classical continuum limit both for an under-damping and an over-damping set of parameters. These are compared with the results obtained in the finite case with $N=600$ bath particles, both with a uniform and Lorentzian sampling of the bath frequencies. The parameters $\omega_\C$ and $a_0$ are chosen so that the Ullersma's spectrum is sampled beyond the cut-off $\alpha$, and the recurrence time is of the order of the characteristic relaxation time $1/\Gamma$.
Finally curves obtained with the adjusted Lorentzian sampling are reported, where the parameter $a_0$ is chosen so that the recurrence time is much longer than $1/\Gamma$.
 
In the under-damping case, for a finite bath and for times shorter than the recurrence time, $C(t)$ exhibits the typical damped oscillating behavior of the continuum limit, apart from a shift in the oscillation frequency $\Omega$. On the other hand, $\sigma_{q,T}^2(t)$ and $\sigma_{p,T}^2(t)$ exhibit in the finite-size case the same dissipative behavior as in the continuum case, with a characteristic time $1/\Gamma$. However, while $\sigma_{p,T}^2(t)$ seems to reach, before the Poincaré recurrence time, the same plateau value $k_\B T$ as in the continuum case, $\sigma_{q,T}^2(t)$ appears to reach a value lower than the one expected, i.e.,~$1/(\beta\Omega_0^2)$. These effects are due to the fact that the frequency shift $\sum_\ell\epsilon_\ell^2/\omega_\ell^2$ is different
from the continuous limit one $\kappa\alpha$, which appears in $\Omega_0^2$. In fact
$\left[\beta\left(\omega_0^2-\sum_\ell\epsilon_\ell^2/\omega_\ell^2\right)\right]^{-1}$ is equal to the plateau value of $\sigma_{q,T}^2(t)$ reached before the recurrence.

In the over-damping case, as the effect of the frequency shift is larger, one observes a larger difference between the continuum and the finite case. In fact, while the continuum limit curves display the typical over-damped behavior without any oscillations, the finite-case curves exhibit the same behavior observed in the under-damping case. Moreover the difference between the plateau values before the recurrence for $\sigma_{q,T}^2(t)$, is also much larger.

It is clear from Figure~\ref{A_C_Q2_P2} that with the Lorenzian sampling one obtains curves that behave more similarly to the continuum ones, for the same bath size and recurrence times, with respect to the uniform case. This holds both for the oscillation frequency of $C(t)$ and the plateau value reached by $\sigma_{q,T}^2(t)$ before the recurrence. One may notice the optimal agreement of the curves obtained with the adjusted Lorentzian distribution with the continuum ones. 

\begin{figure}
 \centering
 \includegraphics[width=\textwidth]{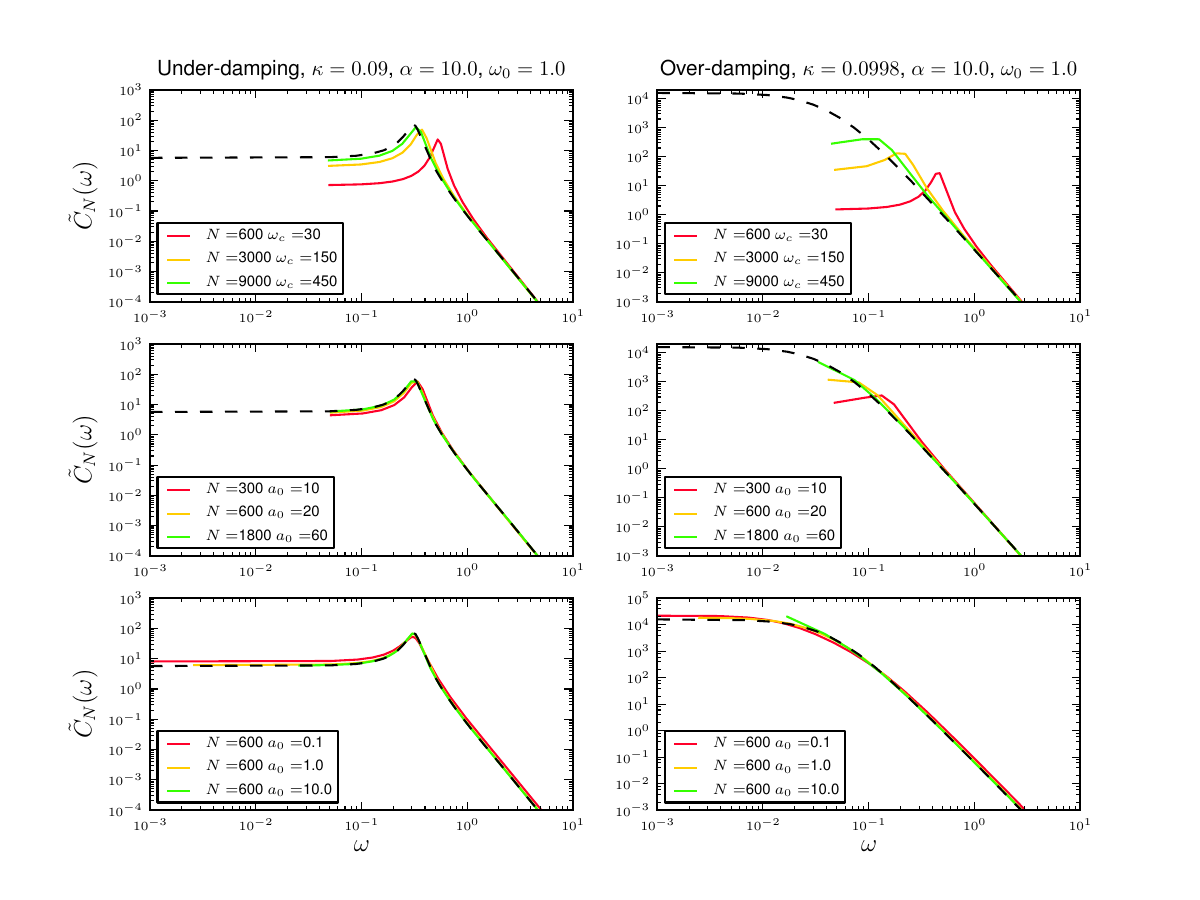}
\caption{Plot of $\tilde{C}_{N}(\omega)$ vs.\ $\omega$
 for different values of the size $N$ of the bath and of the maximal frequancy $\omega_\C$ (center) or the width $a_0$, for a uniform (top), Lorentzian (center) and adjusted Lorentzian (bottom) sampling of bath frequencies. The continuum limit $\tilde{C}(\omega)$ corresponds to the dashed line.}\label{C0_omega}
\end{figure}

The same qualitative behavior of the finite size frequency sampling appear in the Fourier transform of $C(t)$. In Figure~\ref{C0_omega}, with the same parameters of Figure~\ref{A_C_Q2_P2}, one notices that in the under-damping regime
 $\tilde{C}(\omega)$ is characterized by
a peak corresponding to the oscillation frequency $\Omega$. A similar curve characterizes $X_{0\nu}^2/(z_\nu^2\Delta_\nu)$, but the position of the peak is shifted. 
This shift corresponds to the change in the oscillation frequency of $C_N(t)$ with respect
to $C(t)$. In the over-damping case the $N=600$ curve maintains the look
of the under-damping case, while the continuous one looses the peak, then
confirming that in this case there is a worse agreement between the continuous and the
finite cases.

\begin{figure}[ht!]
 \centering
 \includegraphics[width=\textwidth]{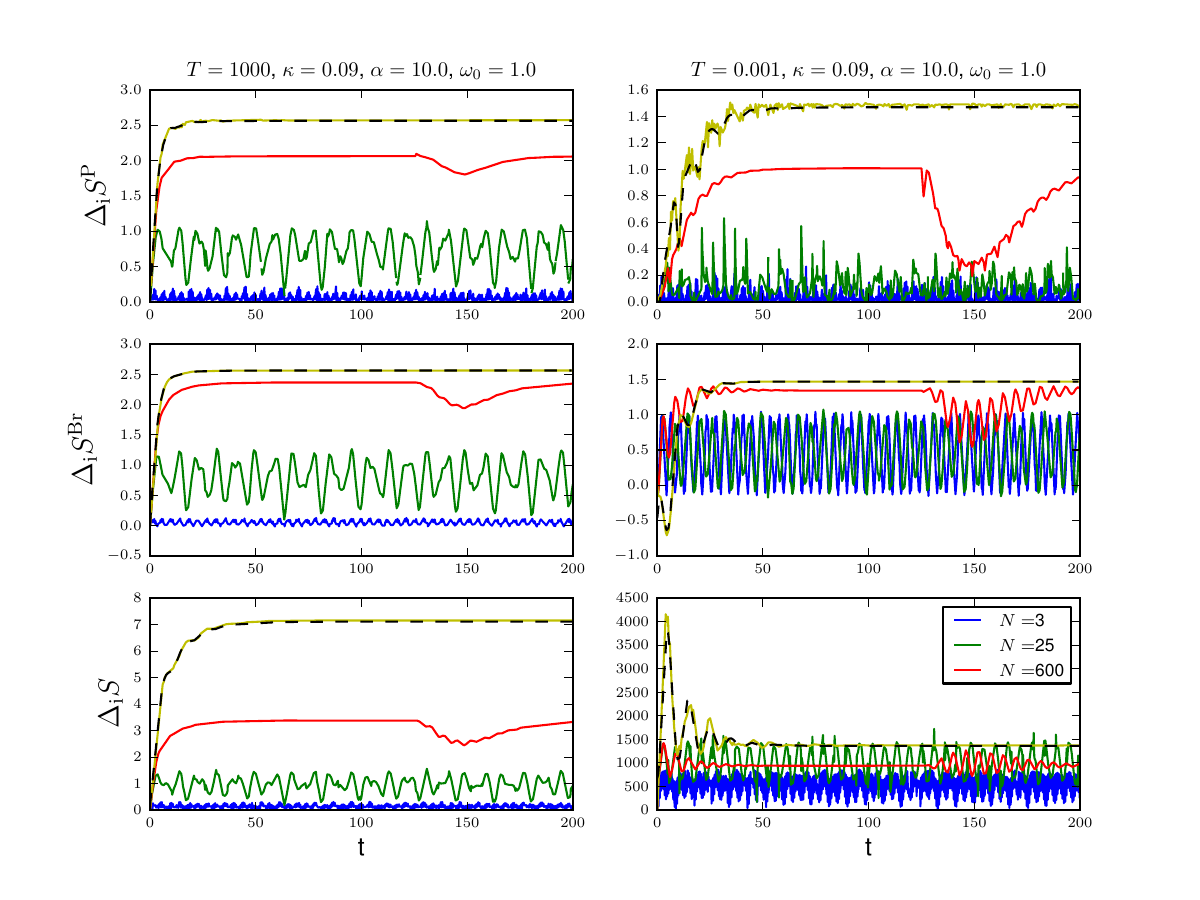}
 \caption{\label{S_vs_N}  Entropy production vs.\ time $t$, according to the the three definitions $\Delta_{\rm i} S^{\P}$, $\Delta_{\rm i} S^{\Br}$ and $\Delta_{\rm i} S$, for a uniform distribution of bath frequencies with a cut-off $\omega_c=30$,  for different sizes $N$: $3$ (blue), $25$ (green), $600$ (red), and for the adjusted Lorentzian distribution with $a_0=0.1$ (yellow) for $N=600$. The dashed black line corresponds to the continuum limit. The initial conditions are those of Figure~\ref{fig:CL_therm_split} in the classical case, apart from the first moments $\average{Q^2(0)}=10$ and $\average{P^2(0)}=10$, while they are those of Figure~\ref{fig:Quan_Br_Esp} in the quantum case.}
\end{figure}

Things improve when  $\omega_\C$ and the size $N$ become larger, keeping the frequency density constant. In this case the peak shifts towards its continuum position in the under-damping case, while in the over-damping case the peak tends to disappear. This improvement is due to the fact that the frequency shift $\sum_\ell\epsilon_\ell^2/\omega_\ell^2$
approaches $\int_0^{\infty}\;\rmd\omega\,\gamma(\omega)/\omega^2=\kappa\alpha$. In fact the difference between these quantities is due to two terms: one given by the difference between the sum $\sum_\ell\epsilon_\ell^2/\omega_\ell^2$ and the integral up to $\omega_{\C}$, which is of order $1/N$ and is negligible for the sizes reported in Figure~\ref{C0_omega}, and one, more relevant, corresponding to the contribution to the integral arising from frequencies larger than $\omega_\C$. This term is proportional to $\kappa$. Thus, in order to maintain the difference between $\sum_\ell\epsilon_\ell^2/\omega_\ell^2$ and $\kappa\alpha$ constant, $\omega_{\C}$ must increase as $\kappa$ increases. In particular for a given set of parameters, which would correspond to over-damping in the continuum limit, one would never obtain over-damping behavior if $\omega_\C$ is too small.

If $\omega_{\C}$ or $a_{0}$ are kept fixed, and $N$ increases, the behavior remains the same, only the recurrence time $t_{\mathrm{P}}$ increases and the smallest frequency $z_{1}$ decreases.

With the Lorentzian sampling of parameters convergence improves both in the under-damping and in the over-damping cases. In fact, by choosing $a_0$ and $N$ so that the recurrence time is of the same order as in the uniform case, the value of the frequency shift is closer to $\kappa\alpha$. This is due to the fact that the highest frequency is much larger. One has to exercise some care in choosing $a_{0}$ neither too large (in order to have long recurrence times) nor too small (in order to avoid too sparse a sampling close to the highest frequency). 

We also show in Figures~\ref{A_C_Q2_P2} and~\ref{C0_omega} the effect of adjusting the  coupling with the highest-frequency oscillator. The behavior of the continuum is optimally matched with the choice $a_0=0.1$ and $N=600$. 

%--------------------------------------------------------------
\subsection{Finite-size entropy production} 

We report in figure~\ref{S_vs_N} the behavior of the entropy production according to the three definitions, i.e., the Poised ($\Delta_{\rm i} S^{\P}$: eq.~(\ref{DS_i_poi})), the Breuer ($\Delta_{\rm i} S^{\Br}$: eq.~(\ref{DS_i_eqbis})) and the ELB ($\Delta_{\rm i} S$: eq.~(\ref{DS_i_esp})), for different values of $N$ in the uniform case and for $N=600$ for the adjusted Lorentzian cases. One notices that in the adjusted Lorentzian case one reaches an almost perfect agreement with the continuum limit already for $N=600$. In the quantum case the expression of the entropy production $\Delta_{\rm i} S$ obtained with the adjusted Lorentzian binning 
does not approximate perfectly the continuum limit. This is due to the poor convergence of the term $\average{Q(t)\eta(t)}$ which is contained in the averaged interaction energy $\average{H_I}$ (\ref{H_int_t}). The same can be observed for $\Delta S_i^{\P}$, due to the noisy behavior of $\dot{A}(t)$ and $\ddot{A}(t)$. At finite sizes the Breuer expression $\Delta_{\rm i} S^\Br$ can assume negative values, whereas both $\Delta_{\rm i} S$ and $\Delta_{\rm i} S^{\P}$ remain positive. However, in the uniform case one obtains a slower convergence with respect to the Lorentzian case, both in the adjusted and in the non-adjusted case (not shown). 

%%%%%%%%%%%%%%%%%%%%%%%%%%%%%%%%%%%%%%%%%%%%%%%%%%%%%%%%%%%%%%%%%%%%%%
\section{Bath entropy}\label{bath_entropy}
%%%%%%%%%%%%%%%%%%%%%%%%%%%%%%%%%%%%%%%%%%%%%%%%%%%%%%%%%%%%%%%%%%%%%%

The bath entropy at time $t$ is given by
\begin{equation} \label{BathEnt}
S_\B(t) =-\tr \rho_\B(t) \ln \rho_\B(t),
\end{equation}
where $\rho_\B(t) =\tr_\sys\rho(t)$ is the reduced bath density matrix.
Since the total density matrix is not a product state $\rho_\sys\otimes\rho_\B$ at times $t>0$, one cannot simply split the total entropy into system entropy plus bath entropy. Thus one introduces the correlation entropy $S_\C$:
\begin{equation}
S_{\mathrm{tot}}=-\tr\rho(t)\ln\rho(t)=S(t)+S_\B(t)+S_\C(t).
\end{equation}
We note that $-S_\C(t)$ is the mutual information between the central oscillator and the bath \cite{Nielsen}.
Since the total entropy is conserved and the initial correlations vanish, one has $S_\C(0)=0$ and, according to this definition,
\begin{equation}\label{S_C}
S_\C(t)=-\Delta S(t)-\Delta S_\B(t).
\end{equation}
We can easily verify that \cite{Esposito}
\begin{equation}\label{S_CasD}
S_\C(t)=-D\left[\rho(t)\Bigg\|\rho_s(t)\prod_r\rho_r(t)\right] \leq 0.
\end{equation}
Thus the correlation entropy is always negative or zero. By comparing this last equation with eq.~(\ref{DS_i_esp}), one finds~\cite{Esposito}
\begin{equation}\label{S_i+S_C}
\Delta_{\rm i} S(t)+S_\C(t)=-\beta \average{\Delta H_{\B}}-\Delta S_{\B}(t)=D[\rho_{\B}(t)\|\rho_{\B}^{\eq}] \geq 0.
\end{equation}
One notices that if the approximation of a bath remaining at equilibrium (ideal bath) were valid, i.e., $\rho_{\B}(t)=\rho_{\B}^{\eq}$, the correlation entropy would be equal  to minus the entropy production $S_\C(t)=-\Delta_{\rm i} S(t)$. In this case the variation of the bath entropy would be equal to the heat flow.

The method to numerically evaluate the bath entropy is detailed in \ref{sec:bathentcalc}. This calculation relies on the fact that the bath density matrix is Gaussian at each time, and therefore is fully characterized by the time-evolving bath covariance matrix (\ref{app:eqtime}). We now turn to the discussion of the results.

\begin{figure}[t]\centering
\includegraphics[width=\textwidth]{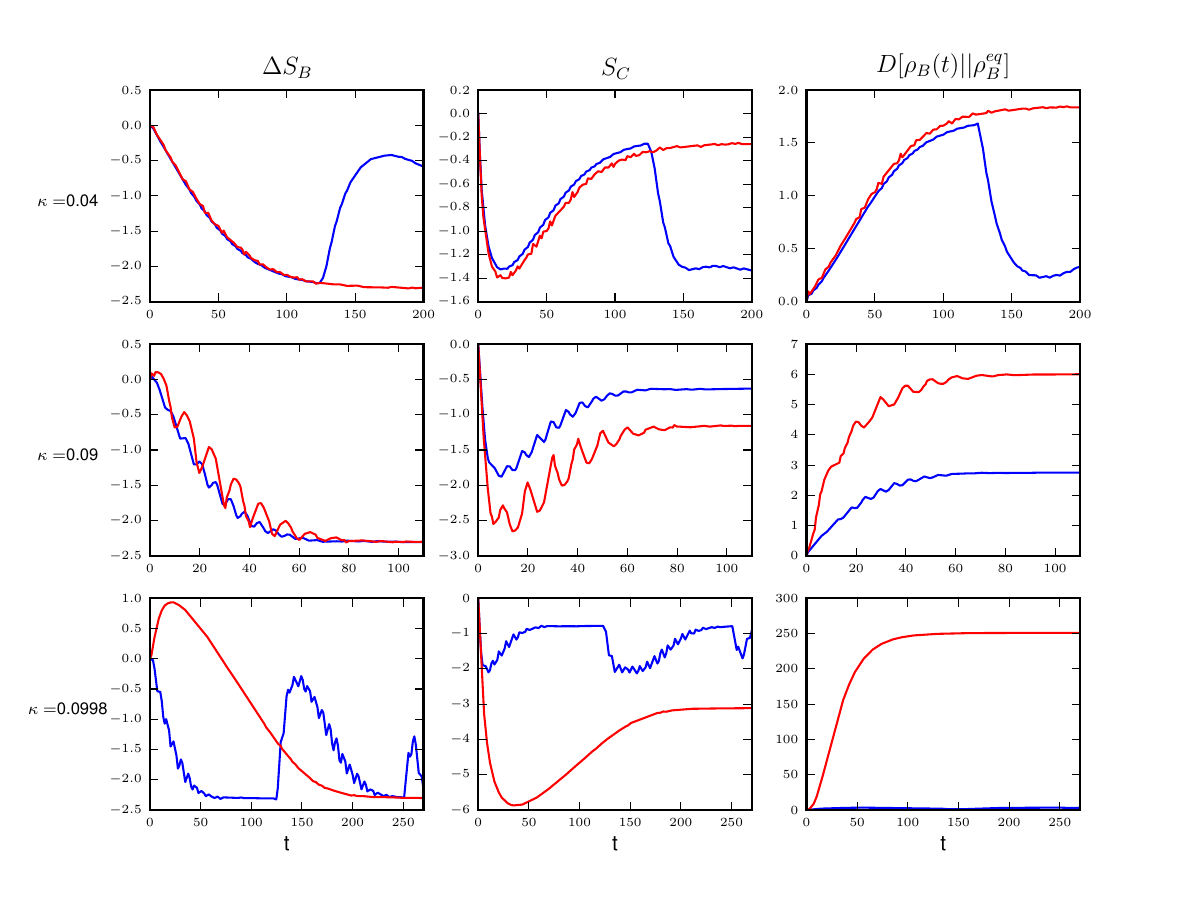}
\caption{\label{Bathent_uni_lor_CL} Entropy change in the bath $\Delta S_B$, correlation entropy $S_C$ and Kulbach-Leibler distance between $\rho_\B(t)$ and $\rho_\B^{\eq}$ in the high temperature $T=1000$ regime, for different values of the coupling $\kappa$, for a uniform sampling of frequencies with $\omega_c=30$ (blue), and for a adjusted Lorentzian one with $a_0=0.1$ (red). The bath size is $N=600$ and other relevant parameters are set as for figure (\ref{S_vs_N}).}
\end{figure}

As observed before for the central oscillator entropy, with an adjusted Lorentzian sampling of frequencies the Poincaré recurrence times are much longer, and one can observe a convergence toward an asymptotic plateau of the different quantities, as shown for the classical case in figure~\ref{Bathent_uni_lor_CL}. In particular the asymptotic value of $\Delta S_\B$ does not depend on the coupling constant $\kappa$, while, in agreement with (\ref{WB_CL}),  it depends on the initial variances of the central system~(see~\ref{app:eqtime}). Interestingly, in the over-damping case $\Delta S_\B$ increases at the beginning, reaching a maximum independently of the initial conditions, before decreasing to the equilibrium value, much as minus the interaction energy (\ref{H_I_t}) does (not shown). The relaxation time is longer than that of the central oscillator. This does not happen in the under-damped regime.

\begin{figure}[t]\centering
\includegraphics[width=\textwidth]{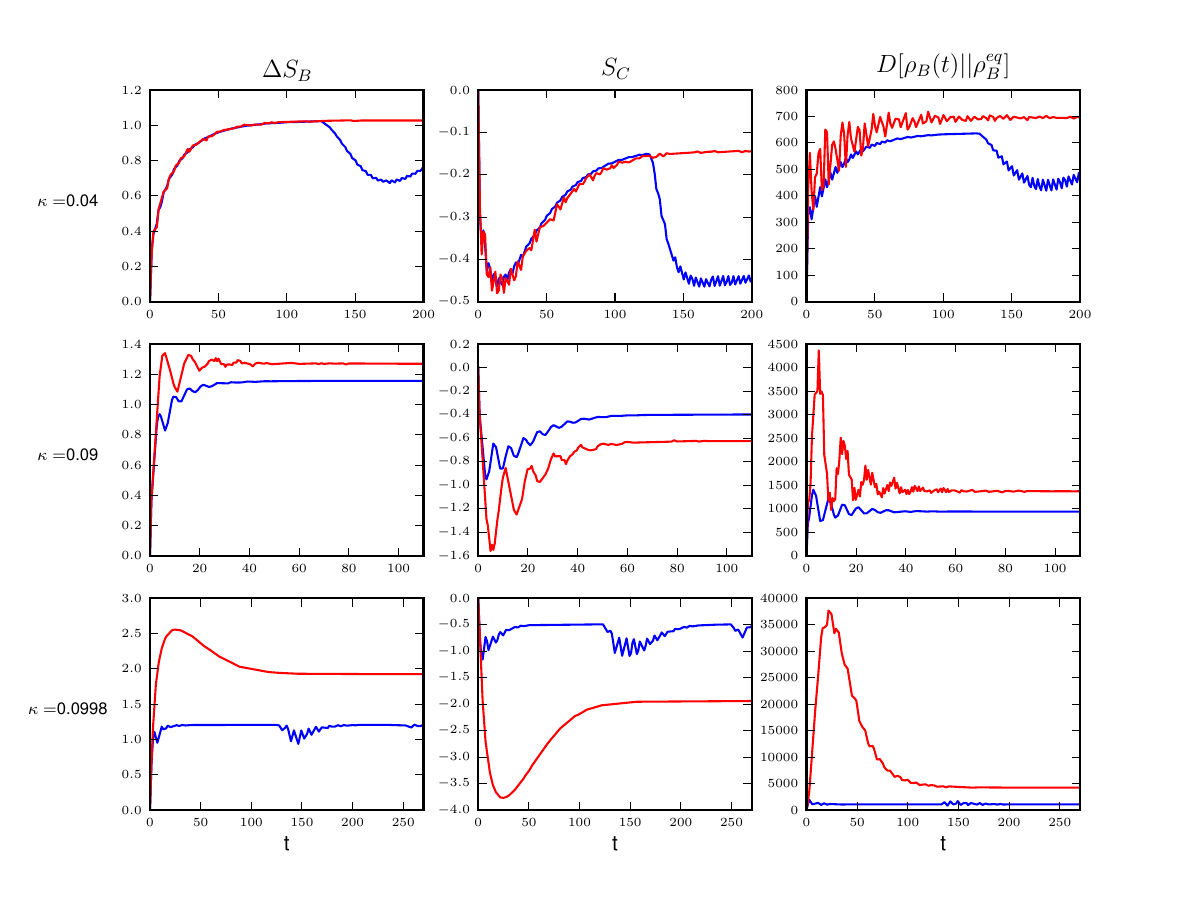}
\caption{\label{Bathent_uni_lor_Q} Same as figure \ref{Bathent_uni_lor_CL} but in the low-temperature case $T=0.001$, with the other parameters set as in figure \ref{fig:Quan_Br_Esp}.}
\end{figure}

Once the bath entropy $\Delta S_\B$ is evaluated, one also gets the correlation entropy $S_\C$ using (\ref{S_C}) and then the distance $D[\rho_\B(t)\|\rho_\B^{\eq}]$ using (\ref{S_i+S_C}). $S_\C$ is negative by definition (\ref{S_CasD}), and its absolute value grows with the coupling as $-\ln (\omega_0^2-\kappa\alpha)$, since $\kappa\alpha$ approaches $\omega_{0}^{2}$, similarly to the entropy of the system $\Delta S$ (\ref{DS_infty_CL}). Like the interaction energy term (\ref{H_I_t}), the asymptotic value of $S_\C$ does not depend on the initial conditions of the central system (data not shown). It turns out instead that the asymptotic value of $S_\C$ vanishes as $\kappa\rightarrow 0$. This is confirmed by the fact that the coupling-independent asymptotic value of $\Delta S_\B$ equals minus the central oscillator entropy change in the limit of vanishing coupling:
\begin{equation}
 \Delta S_\B(\infty)=-\left.\Delta S(\infty)\right|_{\kappa\rightarrow0}=-\left[\ln \frac{T}{\omega_0}-\ln \Delta(0)\right].
\end{equation}

The distance $D(\rho_\B(t)\|\rho_\B^{\eq})$ increases with the coupling as $\kappa\alpha/(\omega_0^2-\kappa\alpha)$, like the negative entropy flow (cf.~equation (\ref{DSe_infty_CL})).
This quantity does not vanish for $\kappa\rightarrow0$, where it equals the entropy production:
\begin{eqnarray}
 \fl\qquad\left.D(\rho_\B(\infty)\|\rho_\B^{\eq})\right|_{\kappa\rightarrow0}&=&\left.\Delta S_\rmi\right|_{\kappa\rightarrow0}=\left[\Delta S(\infty)-\Delta_\rme S(\infty)\right]_{\kappa\rightarrow0}\\
\nonumber&=&\ln \frac{T}{\omega_0}-\ln \Delta(0)-\beta\left[T-\frac{1}{2}(\omega_0^2\average{Q^2(0)}+\average{P^2(0)})\right].
\end{eqnarray}

As a consequence, the bath density matrix operator is always changed and the Kullback-Leibler distance from the density operator at canonical equilibrium becomes larger when increasing the coupling. This suggests that for our model the ideal bath approximation,  namely the assumption $\rho_\B(\infty)\simeq\rho_\B^{\eq}$, which would imply $\Delta S_\B\simeq-\Delta_{\rm e} S=\beta\Delta\average{H_\B}$ (\ref{S_i+S_C}), is not valid even in the thermodynamic limit. One observes in figure~\ref{Bathent_uni_lor_Q} that in the quantum case the asymptotic value of the bath entropy change appears to grow with the coupling, which could be an effect of the entanglement or quantum correlations between the bath and the system. The dependence on the coupling is apparently weaker than that exhibited by the entropy flow. This means that also here the Kullback-Leibler distance between $\rho_\B(t)$ and $\rho_\B^{\eq}$ is relevant and strongly increases with the coupling. Due to the quantum contribution 
in the interaction term, its asymptotic value can
be orders of magnitude larger than the one assumed in the high-temperature limit, in the same way ELB entropy production does.
In the limit of vanishing coupling analogous considerations of the classical case can be made, since the correlation entropy $S_\C$ vanishes. 

%%%%%%%%%%%%%%%%%%%%%%%%%%%%%%%%%%%%%%%%%%%%%%%%%%%%%%%%%%%%%%%%%%%%%%
\section{Conclusions}\label{Conclusions}

In this paper we studied the thermodynamic description of a process of transient relaxation in the QBM model where a central harmonic oscillator initially prepared in a Gaussian nonequilibrium state is bi-linearly coupled with a bath of harmonic oscillators initially prepared at equilibrium. 

We compared two ways of defining entropy production during the ensuing relaxation process of the central oscillator. Both definitions are expressed as the difference between the change is the von Neumann entropy of the system minus an heat divided by the temperature of the reservoir. The `ELB' one is based on defining this heat as minus the energy change in the bath and thus has a straightforward physical interpretation, while the `Poised' one (beyond non-Markovian transients) defines heat in a less transparent way in term of the change in an effective ``mean force'' Hamiltonian. Both expressions are positive by definition but in a general non-Markovian quantum regime they both may exhibit oscillations. However, in the Markovian limit, while the `ELB' may still exhibit oscillations, the `Poised' one becomes a monotonically increasing function of time. The two definitions coincide for vanishing coupling but we have shown that for finite coupling the `ELB' is always larger than
  the `Poised' one. Their difference contains the expectation value of the interaction Hamiltonian and can thus be made arbitrarily large. In the low-temperature limit the contribution due to the quantum corrections in the interaction term can make this difference order of magnitudes larger than in the classical case. Finally, we showed that in the classical over-damped regime the `Poised' one converges to the entropy production defined in stochastic thermodynamics. 

We numerically studied the exact dynamics of our system for a finite number of oscillators in the bath. Using two different samplings of the bath frequencies, a uniform and a Lorentzian ones. In both cases the period of the Poincaré recurrences increases with growing density of bath frequencies but the Lorentzian sampling guarantees a faster convergence to the continuum limit curves as a function of $N$.

Finally, we numerically studied the evolution of the von Neumann entropy of the bath which results from the relaxation process of the central oscillator. This enabled us to calculate the evolution of the system-bath correlation entropy (or minus the mutual information) and the Kullback-Leibler divergence between the bath density matrix at time $t$ and its initial thermal equilibrium form. 
We observed that for a given initial condition of the central oscillator, the asymptotic value of the bath entropy change does not depend on the coupling in the classical limit, while it slightly does in the quantum regime. In the limit of vanishing coupling strength the correlation entropy vanishes, what means that the change in the von Neumann entropy of the bath becomes equal to minus the change in the central system entropy. We also observed that the Kullback-Leibler divergence of the bath density matrix never vanishes, thus indicating that the assumption of an ideal bath which always remains at equilibrium is not satisfied. As expected, this divergence grows significantly with the coupling, as the ELB expression of the entropy production does. 

While our study revealed important features in the QBM model, it also indicates that no definite formulation of a consistent thermodynamics of out-of-equilibrium quantum systems in presence of non-vanishing coupling with the bath is yet available.   

%%%%%%%%%%%%%%%%%%%%%%%%%%%%%%%%%%%%%%%%%%%%%%%%%%%%%%%%%%%%%%%%%%%%%%
\ack

This research was supported by the Dottorato in Fisica Fondamentale e Applicata, Università ``Federico~II'', by PRIN 2009PYYZM5, by the National Research Fund Luxembourg in the frame of project FNR/A11/02, and by the ESF networking program ``Exploring the physics of small devices".

%%%%%%%%%%%%%%%%%%%%%%%%%%%%%%%%%%%%%%%%%%%%%%%%%%%%%%%%%%%%%%%%%%%%%%
\appendix
%%%%%%%%%%%%%%%%%%%%%%%%%%%%%%%%%%%%%%%%%%%%%%%%%%%%%%%%%%%%%%%%%%%%%%
\section{Ullersma's solution}\label{Usol}

Solution (\ref{H-solutions}) is obtained by first finding a matrix transformation into new conjugate operators $\{Q'_\mu,P'_\mu\}$ which diagonalize the Hamiltonian into a set of $N+1$ normal harmonic oscillators, by then writing the Heisenberg solutions in that basis, and by finally transforming
back to the original operators.

The functions $A_{\mu\nu}(t)$ can be expressed in terms of the function
\begin{equation}\label{g_z}
 g(z)=z^2-\omega_0^2-\sum_{i=1}^N\frac{\epsilon^2_i}{z^2-\omega_i^2},
\end{equation}
whose zeros $z_\nu$, $\nu=0,\ldots,N$, are the normal frequencies of the harmonic oscillators in the new basis.
We have in fact
\begin{equation}
 A_{\mu\nu}(t)=\sum_{\rho=0}^N X_{\mu\rho}X_{\nu\rho}\frac{\sin(z_\rho t)}{z_\rho},
\end{equation}
where the elements $X_{\mu\nu}$ of the transformation matrix are given by
\begin{eqnarray}
X_{0\nu}=\left[\frac{1}{2z}\left.\frac{\rmd g(z)}{\rmd z}\right|_{z=z_\nu}\right]^{-1/2},\qquad \nu=0,\ldots,N;\\
X_{i\nu}=\frac{\epsilon_i}{z_\nu^2-\omega_i^2}X_{0\nu},\qquad i=1,\ldots,N;\quad \nu=0,\ldots,N.
\end{eqnarray}

%%%%%%%%%%%%%%%%%%%%%%%%%%%%%%%%%%%%%%%%%%%
\section{Quantum Langevin Equation}\label{QLE}

In order to obtain the QLE, we first exploit the explicit solution to write down the equations of motion for the position operators in the bath as integro-differential equations involving the position operator of the central oscillator:
\begin{equation}\label{Qi_t_formal}
Q_i(t)=Q_i(0)\cos(\omega_it)+\frac{P_i(0)}{\omega_i}\sin(\omega_it)-\frac{\epsilon_i}{\omega_i}\int_0^t\rmd s\;\sin[\omega_i(t-s)]Q(s).
\end{equation}
Then the central oscillator satisfies the following Quantum Langevin Equation (QLE) which we express in a matrix representation:
\begin{equation}\label{Lang-Matr}
\dot{{\bm z}}(t)+{\bm H}\ast{\bm z}(t)=-{\bm \eta}(t)-{\bm F}(t).
\end{equation}
In this expression, $\ast$ represents the time convolution, $\bm{H}(t)$ is given by
\begin{equation}
{\bm H}(t)=\left[\begin{array}{c c}
0 & -\delta(t)\\
\Omega_0^2\delta(t) & K(t)
\end{array}\right];
\end{equation}
where $K(t)$ is the damping kernel and ${\bf F}(t)=(0,K(t)Q(0))$ is the forcing term, which is responsible in the continuum limit for a fast slip of the initial conditions (see section~\ref{therm_limit}). 

The solution of the differential equation (\ref{Lang-Matr}) can be easily obtained by taking the Laplace transform and then transforming back. One obtains
expression (\ref{Lang_sol}) where both terms contain the matrix propagator $\bm{\Phi}(t)$
depending on the damping kernel $K(t)$ via the propagator $A(t)$. The Laplace transform of $A(t)$ is given by
\begin{equation}\label{Lap_A}
\widehat{A}(s)=\frac{1}{s^2+s\widehat{K}(s)+\Omega_0^2}.
\end{equation}

In~\ref{app:equivalence} we show that the Ullersma solution (\ref{H-solutions}) and the Fleming one (\ref{Qi_t_formal},\ref{Lang_sol}) are equivalent.

%%%%%%%%%%%%%%%%%%%%%%%%%%%%%%%%%%%%%%%%%%%%%%
\section{Master Equation matrices}\label{ME-Matr}

The \emph{pseudo-Hamiltonian} ${\bm{\mathcal{H}}}(t)$ and \emph{diffusion} ${\bm D}(t)$ matrices \cite{Fleming} are defined as:
\begin{eqnarray}
&&\hspace{-1cm} \bm{\mathcal{H}}(t) 
\equiv \left[\begin{array}{cc} 0 & -1 \\ \Omega_\mathrm{R}^2(t) & 2\Gamma(t) \end{array} \right] 
=-\dot{\bm\Phi}(t){\bm\Phi}^{-1}(t) ,\\\label{eq:diffusion}
&&\hspace{-1cm} \bm{D}(t) 
\equiv \left[\begin{array}{cc} 0 & -\frac{1}{2}D_{qp}(t) \\ -\frac{1}{2}D_{qp}(t) & D_{pp}(t) \end{array} \right] 
= \frac{1}{2}\left[\bm{\mathcal{H}}(t){\bm\sigma_T}(t)+{\bm \sigma}_T(t)\bm{\mathcal{H}}^{\mathsf{T}}(t)+\dot{\bm{\sigma}}_T(t)\right], 
\end{eqnarray}
where ${\bm\Phi}(t)$ is the matrix propagator (\ref{M_prop}) and the thermal covariance matrix $\bm{\sigma}_T(t)$ is defined by
\begin{equation}\label{thermal_M}
\hspace{-2cm} \bm{\sigma}_T(t) = \left[\begin{array}{cc} \sigma^2_{q,T} & C_{qp,T} \\ C_{qp,T} & \sigma^2_{p,T} \end{array} \right]
= \int_0^t\rmd\tau\int_0^t\rmd\tau'\;{\bm \Phi}(t-\tau) \left[\begin{array}{cc} 0 & 0 \\ 0 & \nu(\tau-\tau') \end{array} \right] {\bm\Phi} ^{\mathsf{T}}(t-\tau').
\end{equation}
where $\nu$ is the noise kernel.

%%%%%%%%%%%%%%%%%%%%%%%%%%%%%%%%%%%%%%%%%%%%%%
\section{Equivalence of the Ullersma and Fleming solutions}\label{app:equivalence}

The convolution with the noise in (\ref{Lang_sol}) corresponds to the sum over the bath operators in (\ref{H-solutions}) and the expression for bath operators in (\ref{H-solutions}) correspond exactly to that in (\ref{Qi_t_formal}). This equivalence is recovered thanks to the following equation
relating the quantities $A_{i0}(t)$ to the propagator $A(t)$,  and to the equation relating quantities $A_{ij}(t)$ to $A_{i0}(t)$, where $i$ and $j$ are bath indices:
\begin{eqnarray}\label{A_i0-A}
&&\ddot{A}_{i0}(t)+\omega_i^2A_{i0}(t)=-\epsilon_iA(t);\\
&&\ddot{A}_{ij}(t)+\omega_j^2A_{ij}(t)=-\epsilon_jA_{i0}(t).
\end{eqnarray}
Given the initial conditions $\dot{A}_{i0}(t)=A_{i0}(t)=A_{ij}(t)=0$ and $\dot{A}_{ij}(t)=\delta_{ij}$ these equations imply that 
\begin{eqnarray}\label{A_i0(A)}
&&A_{i0}(t)=-\epsilon_i\int_0^td\tau A(\tau)\frac{\sin[\omega_i(t-\tau)]}{\omega_i};\\\label{A_ij(A_i0)}
&&A_{ij}(t)=\frac{\sin(\omega_jt)}{\omega_j}\delta_{ij}-\epsilon_j\int_0^td\tau A_{i0}(\tau)\frac{\sin[\omega_j(t-\tau)]}{\omega_i}.
\end{eqnarray} 
%%%%%%%%%%%%%%%%%%%%%%%%%%%%%%%%%%%%%%%%%%%%%%
\section{Time scales of the propagator}\label{Timescales}

\begin{figure}[h]
\centering
\includegraphics[width=0.7\textwidth]{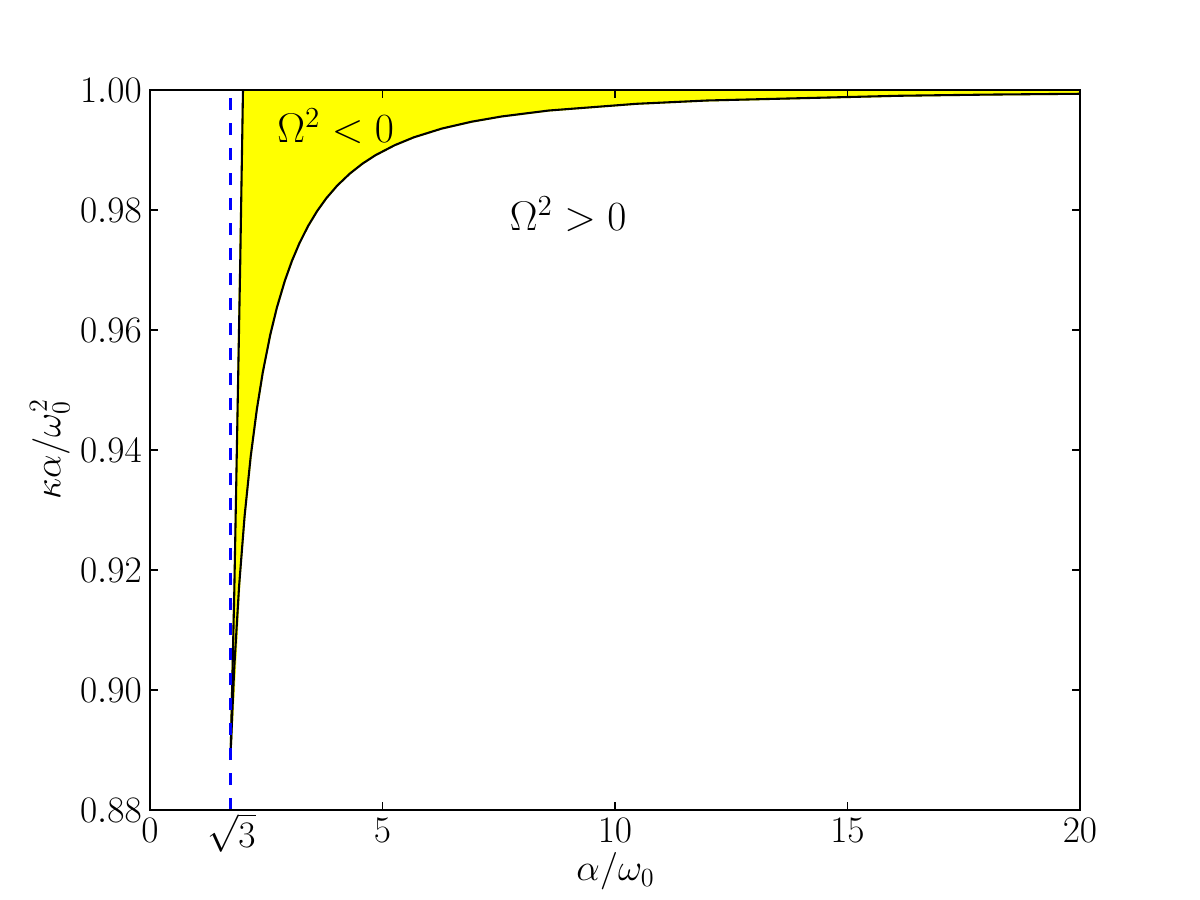}
\caption{Transition between real (white) and imaginary (yellow) $\Omega$. In the large cut-off limit ($\ref{high-cut-off}$) it corresponds to the transition between under-damping $\Gamma<\Omega_0$ and over-damping $\Gamma>\Omega_0$. For $\alpha/\omega_0\le \sqrt{3}$ (on the left of the dotted line) one always has $\Omega^{2}>0$.}\label{Transition} 
\end{figure}
Time scales $\Omega$, $\Gamma$ and $\lambda$ are obtained by solving the following equations in which the bare central oscillator frequency $\omega_0$, the coupling $\kappa$ and the cut-off $\alpha$ appear:
\begin{eqnarray}\label{eq:lambda}
\lambda+2\Gamma=\alpha \ \ , \ \
\Omega^2+\Gamma^2+2\Gamma\lambda=\omega_{0}^{2} \ \ , \ \ 
\left(\Omega^2+\Gamma^2\right)\frac{\lambda}{\alpha}=\omega_0^2-\kappa\alpha.                 
\end{eqnarray}
In the large cut-off limit (\ref{high-cut-off}) can then approximate (\ref{eq:lambda}) by
\begin{eqnarray}
\lambda\simeq\alpha \ \ , \ \ \Gamma \simeq \kappa/2 \ \ , \ \ \Omega_0^{2} \simeq \Gamma^2+\Omega^2,
\end{eqnarray}

The pseudo-Hamiltonian $\bm{\mathcal{H}}(t)$ in (\ref{Wig_me}), at~$\Order{1/\alpha}$ and for $t\gg1/\alpha$, becomes equal to the the time-independent matrix
\begin{equation}
 \bm{\mathcal{H}}_{\loc} =\left[\begin{array}{cc}
 0 & -1\\
\Omega_0^2 & 2\Gamma
\end{array}\right]
\end{equation}
which is characteristic of a Ornstein-Uhlenbeck process.
Since $\Omega_0$ is real, the transition between real and imaginary $\Omega$ corresponds to the transition between the
under-damped ($\Gamma<\Omega_0$) and the over-damped regimes ($\Gamma>\Omega_0$) (see figure \ref{Transition}).

%%%%%%%%%%%%%%%%%%%%%%%%%%%%%%%%%%%%%%%%%%%%%
\section{Covariance matrix} \label{num-cov-matr}

In order to evaluate the covariance matrix of the central oscillator, one
evaluates the first and second moments for position and momentum Heisenberg operators. From the Heisenberg solutions (\ref{H-solutions}), one obtains by averaging over initial conditions (\ref{factor_ic}) (\ref{Gauss-Wig}) (\ref{bath_ic}) the following expressions:
\begin{eqnarray}\label{mom}
&&\average{Q(t)} =\dot{A}(t)\average{Q(0)}+A(t)\average{P(0)};\\
&&\average{P(t)} =\ddot{A}(t)\average{Q(0)}+\dot{A}(t)\average{P(0)};\\
&&\sigma_q^2(t) =\average{Q^2(t)}-\average{Q(t)}^2\nonumber\\
 &&\qquad{}=\dot{A}^2(t)\sigma_q^2(0)+2\dot{A}(t)A(t)C_{qp}(0)+A^2(t)\sigma_p^2(0)+\sigma^2_{q,T}(t);\\
&&\sigma_p^2(t) =\average{P^2(t)}-\average{P(t)}^2\nonumber\\
 &&\qquad=\ddot{A}^2(t)\sigma_q^2(0)+2\ddot{A}(t)\dot{A}(t)C_{qp}(0)+\dot{A}^2(t)\sigma_p^2(0)+\sigma^2_{p,T}(t);\\ 
&&C_{qp}(t) =\frac{1}{2}\left<\{Q(t)-\average{Q(t)},P(t)-\average{P(t)}\}\right>\nonumber\\
 &&\qquad{}=\frac{1}{2}\frac{\rmd}{\rmd t}\sigma_q^2(t).
\end{eqnarray}
The thermal parts of the covariance matrix for a finite bath have the form
\begin{eqnarray}\label{sigma-therm-discr} 
&&\sigma^2_{q,T}(t) =\sum_{\ell=1}^N\left[\dot{A}_{\ell 0}^2(t)/\omega_\ell^2+A_{\ell 0}^2(t)\right]E(\omega_\ell,T);\\
&&\sigma^2_{p,T}(t) =\sum_{\ell=1}^N\left[\ddot{A}_{\ell 0}^2(t)/\omega_\ell^2+\dot{A}_{\ell 0}^2(t)\right]E(\omega_\ell,T).\label{sigma-therm-discr-p2}
\end{eqnarray}
The latter can be also generally written in integral form as shown in (\ref{CM_sist}). By combining equations (\ref{mom}),(\ref{sigma-therm-discr}) and  one easily obtains also the second moments of the momentum and position operators.

With the continuous bath with the Ullersma coupling strength (\ref{U_strenght}) and the high cut-off limit (\ref{high-cut-off}), by inserting (\ref{A_local}) in the equations (\ref{CM_sist}), one obtains
\begin{eqnarray}\label{sigma-therm}
\fl\sigma^2_{q,T}(t) =\{1+a_\loc^2(t)\}\average{Q^2}_{\eq}+A^2_\loc(t)\average{P^2}_{\eq}+2\left[A_\loc(t)\dot{C}(t)-a_\loc(t)C(t)\right];\\
\fl\sigma^2_{p,T}(t) =\Omega_0^4A^2_\loc(t)\average{Q^2}_{\eq}+\{1-\dot{A}^2_\loc(t)\}\average{P^2}_{\eq}\nonumber\\
 {}\qquad{}+2\left[\Omega_0^2A_\loc(t)\dot{C}(t)+\dot{A}_\loc(t)\ddot{C}(t)\right];\label{sigma-therm-P2}
\end{eqnarray}
where $a_\loc(t)=\dot{A}_\loc(t)+2\Gamma A_\loc(t)$, and the position equilibrium correlation function (\ref{C_t}) is given by
\begin{eqnarray}\label{C_t}
 \fl\qquad C(t) =\frac{1}{2}\average{\{Q(t),Q(0)\}}_{\eq}=\int_0^\infty \rmd\omega\;\frac{\gamma(\omega)}{\omega^2}E(\omega,T)\left|\int_0^\infty \rmd t'\;A_\loc(t')\rme^{\rmi\omega t'}\right|^2;\\
\fl\qquad\int_0^\infty \rmd t'A_\loc(t')\,\rme^{\rmi\omega t'} =\frac{1}{\left(\Gamma-\rmi \omega\right)^{2}+\Omega^{2}}.
\end{eqnarray}
When calculated with $A_\loc(t)$, it differs from the exact one by corrections of~$\Order{1/\alpha}$:
\begin{eqnarray}\label{C_l}\fl
 C(t) =\frac{1}{\beta\Omega_0^2}a_\loc(t)+\frac{\hbar}{2\pi\Omega}\Im\left\{\rme^{-(\Gamma+i\Omega)t}\left[\psi(1+(\Gamma+i\Omega)\tau_\beta)\right.\right.\nonumber\\
\left.\left.{}-\psi(1+(\Gamma-i\Omega)\tau_\beta)\right]\right\}+C_{\alpha,\tau_\beta}(t),\\
\fl
C_{\alpha,\tau_\beta}(t) =\kappa\hbar\left[\frac{1}{2\alpha^2}\frac{\cot(\pi\alpha\tau_\beta)\rme^{-\alpha t}}{((1+\frac{\Gamma}{\alpha})^2+(\frac{\Omega}{\alpha})^2)((1-\frac{\Gamma}{\alpha})^2+(\frac{\Omega}{\alpha})^2)}\right.\nonumber\\
\left.{}-\frac{1}{\pi}\sum_{\ell=1}^{\infty}\frac{(\alpha\tau_\beta)^2}{(\alpha\tau_\beta)^2-\ell^2}\frac{ \ell\tau_\beta^2\rme^{-\ell t/\tau_\beta}}{((\Omega\tau_\beta)^2+(\ell+\Gamma\tau_\beta)^2)((\Omega\tau_\beta)^2+(\ell-\Gamma\tau_\beta)^2)}\right].
\end{eqnarray}
In this expression, $\psi(z)=\rmd \ln \Gamma_{\mathrm{E}}(z)/\rmd z$ is the digamma function, and $\tau_\beta$ was defined in \ref{eq:taubeta}. The last term contains the so called thermal transients, which vanish slowly in the low-temperature limit. By discarding terms of~$\Order{1/\alpha^2}$, in the quantum limit $\alpha\tau_\beta\gg1$ it can be approximated for $t\gg1/\alpha$ by the series \cite{Haake}
\begin{equation}\label{C_alpha_approx}
 C_{\alpha,\tau_\beta}(t)\simeq\frac{\hbar}{\pi\Omega}\Im\sum_{\ell=1}^\infty\frac{\ell\rme^{-\ell t/\tau_\beta}}{(\Gamma+i\Omega)^2\tau_\beta^2-\ell^2}.
\end{equation}
One should remark however that its second time derivative diverges at $t=0$, and that other terms should be taken in account in order to remove this divergence. We truncate this sum to 50 terms, what guarantees a good description for $t\gg1/\alpha$. One has however to take into account the fact that our approximations do not describe well the behavior for $t\leq1/\alpha$.

The equilibrium second moments one gets from the equilibrium correlation function (\ref{C_t}) at $t=0$ are given by \cite{Haake} 
\begin{eqnarray}\label{Q2-eq}\fl
 &&\average{Q^2}_{\eq} =C(0)=\frac{T}{\Omega_0^2}+\frac{\hbar}{\pi\Omega}\Im\psi(1+(\Gamma+i\Omega)\tau_\beta);\\
 \label{P2-eq}
 &&\average{P^2}_{\eq} =-\ddot{C}(0)=T+\frac{2}{\pi}\hbar\Gamma \Re\left(\ln\alpha/\nu-\psi(1+(\Gamma+i\Omega)\tau_\beta)\right)\nonumber\\
&&\qquad\qquad{}+\frac{\hbar(\Omega^2-\Gamma^2)}{\pi\Omega}\Im\psi(1+(\Gamma+i\Omega)\tau_\beta).
\end{eqnarray}
Here the average is carried over the equilibrium density matrix $\rho^{\eq}$ (\ref{rho_st_eq}). This time, in order to correctly evaluate $\average{P^2}_{\eq}$ to~$\Order{1/\alpha}$, one has to consider all the terms contained in $C_{\alpha,\tau_\beta}(t)$. One should keep in mind that $\average{P^2}_{\eq}$ contains a contribution $\ln\alpha/\nu$, what explains the necessity of introducing a high-frequency cut-off. %This is not present in the averaged square position at equilibrium.

%%%%%%%%%%%%%%%%%%%%%%%%%%%%%%%%%%%%%%%%%%%%%%%%%%%%%%%%%%%%%%%%%%%
\section{Effect of initial slips}\label{sec:slips}

As we have recalled in section (\ref{therm_limit}), slips in the averaged momentum operator and in the correlation matrix are produced by the
kick-like force term $F(t)$ in the QLE ~(\ref{Lang-Matr}), acting during an inital time interval of duration $\sim 1/\alpha\simeq 1/\lambda$. The local propagator (\ref{A_local}) actually contains such slips from $t=0^{+}$, so that it is correct
 apart from corrections of~$\Order{1/\lambda}$ only for $t\gg 1/\lambda$ (see eq.~\ref{ddA_slip}), namely when the kick vanishes.
In fact, while $A(0)$ contains  corrections of the kind $(1/\lambda^2)\rme^{-\lambda t}$ with respect to $A_\loc(0)$, its second
time derivative contains a term $\rme^{-\lambda t}$, so that it is negligible only for $t\gg 1/\lambda$.
With the local propagator (\ref{A_local}) we are not going to consider the detail of the evolution
in the initial time interval of duration $1/\alpha$, which is considered to be much shorter than the
other time scales in the large cut-off limit.

The initial slips correspond to a fast shift of the initial conditions:
\begin{eqnarray}\label{i-slips}
 &&\average{P(0)}\rightarrow-2\Gamma\average{Q(0)}+\average{P(0)};\nonumber\\
&&\average{P^2(0)}\rightarrow4\Gamma^2\average{Q^2(0)}-2\Gamma\average{\frac{\{Q(0),P(0)\}}{2}}+\average{P^2(0)};\\
&&\average{\frac{\{Q(0),P(0)\}}{2}}\rightarrow-2\Gamma\average{Q^2(0)}+\average{\frac{\{Q(0),P(0)\}}{2}}.\nonumber
\end{eqnarray}
Let us now discuss the effect of the initial slips (\ref{i-slips}) on the Breuer (\ref{DS_i_eq}) and ELB (\ref{DS_i_esp}) entropy definitions. They appear as a nonvanishing value for $\lim_{t\to 0^{+}}\Delta_{\rm e} S(t)=\Delta_{\rm e} S(0^{+})$ and $\lim_{t\to 0^{+}}\Delta S(t)=\Delta S(0^{+})$.

For the Breuer entropy flow one has, by using the expressions of the moments reported in~(\ref{mom}):
\begin{equation}
 \Delta_{\rm e} S^{\Br}(0^{+})=\beta\left[\frac{1}{M_{\eff}}(2\Gamma\average{Q^2(0)}-\Gamma C_{qp}(0))\right],
\end{equation}
which is due to the shift on $\average{P^2(0)}$. For the entropy change one has
\begin{eqnarray}
 &&\Delta S(0^{+}) =(\Delta_{is}(0)+1)\ln(\Delta_{is}(0)+1)-\Delta_\mathrm{is}\ln\Delta_\mathrm{is}\nonumber\\
&&\qquad\qquad-(\Delta(0)+1)\ln(\Delta(0)+1)-\Delta(0)\ln\Delta(0);\\
&&\Delta_\mathrm{is} =(\sigma_q^2(0)\sigma_p^2(0)-C_{qp}(0)(C_{qp}(0)-2\Gamma\sigma_q^2(0)))^{\frac{1}{2}}-\frac{1}{2},
\end{eqnarray}
where $\Delta(0)$ is given in (\ref{Sist_entr_gauss}). $\Delta_\mathrm{is}>\Delta(0)\geq0$ satisfies the Heisenberg principle, and therefore $\Delta S(0^{+})$ is always positive.

For the ELB entropy flow one has
\begin{equation}
 \Delta_{\rm e} S(0^{+})=\beta\left[2\Gamma\average{Q^2(0)}-\Gamma C_{qp}(0)-\kappa\alpha\average {Q^2(0)}\right].
\end{equation}
 Here the sign is determined by the last term, which is generally larger than the first one, due
to the large value assumed by $\alpha$. 

In principle one should also consider a slip term $\frac{4}{\pi}\hbar\Gamma\psi(1+\lambda\tau_\beta)$ in $\average{P^2(0)}$, because we used the approximate
 formula (\ref{C_alpha_approx}) to calculate (\ref{sigma-therm-P2}). However, the effect of neglecting this term as well as the effect of the truncation of the sum in (\ref{C_alpha_approx}) are negligible
 compared to the slips considered above.

%%%%%%%%%%%%%%%%%%%%%%%%%%%%%%%%%%%%%%%%%%%%%%%%%%%%%%%%%%%%%%%%%%%%%
\section{Calculation of the Poised entropy production}\label{app:breuer}

In order to evaluate the Poised entropy production $\Delta_{\rm i} S^{\P}$ in eq.(\ref{DS_i_poi}), we have to find the poised density matrix $\rho^*_{\sys}(t)$. Once it is known, on can evaluate $\Delta_{\rm i} S^{\P}$ as
\begin{eqnarray}
\Delta_{\rm i} S^{\P}=\Delta S-\Delta_{\rm e} S^{\P}=\Delta S-\Tr_{\sys}\left(\rho_{\sys}(0)-\rho_{\sys}(t)\right)\ln\rho_{\sys}^*(t).
\end{eqnarray}

In order to evaluate $\rho^*_\sys(t)$, we rewrite equation $V(t)\rho_{\sys}^*(t)=\rho_{\sys}^*(t)$ in the Fourier transform space associated with the corresponding Wigner $W_\sys^*(q,p,t)$. Using eq.(\ref{W_S_Fourier}) we get
\begin{equation}\label{eq:W*}
 \widetilde{W}_{\sys}^*({\bm\Phi} ^{\mathsf{T}}(t)\bm{k},t)\,\rme^{-\frac{1}{2}\bm {k} ^{\mathsf{T}}\bm {\sigma}_T(t)\bm {k}}=\widetilde{W}_{\sys}^*(\bm{k},t).
\end{equation} 
Since we only consider initial Gaussian distributions, we have seen that the solution remains Gaussian at any time. Therefore, to solve (\ref{eq:W*}), we look for solutions of the form
\begin{equation}\label{W*_gauss}
 \widetilde{W}_{\sys}^*(\bm{k},t)=\rme^{-\frac{1}{2}\bm {k} ^{\mathsf{T}}\bm {\sigma}^*(t)\bm {k}-\rmi\bm {k} ^{\mathsf{T}}\bm{z}^*(t)},
\end{equation}
where $\bm {\sigma}^*(t)$ is a symmetric $2\times2$ covariance matrix:
\begin{equation}
 \bm {\sigma}^*(t)=\left[\begin{array}{cc}
        {\sigma^{*}_{q}}^2(t) & C^{*}_{qp}(t) \\
	C^{*}_{qp}(t) & {\sigma^{*}_{p}}^2(t)
       \end{array} \right],
\end{equation}
and the vector $\bm{z}^*(t)$ contains the first moments $q^*(t)$ and $p^*(t)$.
By using expression~(\ref{W*_gauss}) in equation~(\ref{eq:W*}) one straightforwardly finds the relation between the covariance matrices and the first moments:
\begin{eqnarray}\label{eq:sigma*1}
 &&{\bm\Phi}(t)\bm {\sigma}^*(t){\bm\Phi} ^{\mathsf{T}}(t)+\bm {\sigma}_T(t)=\bm {\sigma}^*(t);\\\label{eq:z*}
&&{\bm\Phi}(t)\bm{z}^*(t)=\bm{z}^*(t).
\end{eqnarray}
From equation (\ref{eq:z*}) one finds $\bm{z}^*(t)=0$ at all times. Equation~(\ref{eq:sigma*1}) for the covariance matrix is equivalent to a system of three equations. One gets 
\begin{eqnarray}\label{solutions:cov_star1}
 &&{\sigma^{*}_{q}}^{2}(t)=\frac{\sigma^{2}_{q,T}(t)S_{11}(t)+C_{qp,T}(t)S_{12}(t)+\sigma^{2}_{p,T}(t)S_{13}(t)}{D_{11}(t)D_{12}(t)D_{21}(t)};\\ \nonumber
&&C^{*}_{qp}(t)=\frac{C_{qp,T}(t)S_{22}(t)+\sigma^{2}_{q,T}(t)S_{21}(t)+\sigma^{2}_{p,T}(t)S_{23}(t)}{D_{12}(t)D_{21}(t)};\\\nonumber
 &&{\sigma^{*}_{p}}^{2}(t)=\frac{\sigma^{2}_{p,T}(t)S_{33}(t)+C_{qp,T}(t)S_{32}(t)+\sigma^{2}_{q,T}(t)S_{31}(t)}{D_{11}(t)D_{12}(t)D_{21}(t)}.\nonumber
\end{eqnarray}
Here
\begin{eqnarray}
 &&S_{11}=S_{33}=1-2\dot{A}^2+\dot{A}^4-A\ddot{A}-\dot{A}^2A\ddot{A};\\\nonumber
&&S_{12}=-2\dot{A}A(1-\dot{A}^2+A\ddot{A});\\\nonumber
&&S_{13}=-A^2(1+\dot{A}^2-A\ddot{A});\\\nonumber
&&S_{21}=\dot{A}\ddot{A};\\\nonumber
&&S_{22}=1-\dot{A}^2-A\ddot{A};\\\nonumber
&&S_{23}=A\dot{A};\\\nonumber
&&S_{31}=-\ddot{A}^2(1+\dot{A}^2-A\ddot{A});\\\nonumber
&&S_{32}=-2\dot{A}\ddot{A}(1-\dot{A}^2+A\ddot{A});\\\nonumber
&&D_{11}=1-\dot{A}^2+A\ddot{A};\\\nonumber
&&D_{12}=1-2\dot{A}+\dot{A}^2-A\ddot{A};\\\nonumber
&&D_{21}=1+2\dot{A}+\dot{A}^2-A\ddot{A}.
\end{eqnarray}
Thus the elements of the matrix $\bm {\sigma}^*(t)$ are expressed as linear combinations of thermal covariance elements, whose coefficients 
are functions of the propagator matrix elements. Expressions (\ref{solutions:cov_star1}) are valid both in the finite-size case and in the thermodynamic limit. In order to get the poised covariance matrix at $t=0$ one has to evaluate the $t\rightarrow0^+$ limit of (\ref{solutions:cov_star1}).

%%%%%%%%%%%%%%%%%%%%%%%%%%%%%%%%%%%%%%%%%%%%%%%%%%%%%%%%%%%%%%%%%%%%%
\section{Liouvillian operator of the adjoint dynamics}\label{sec:TR_dynamics}

In the adjoint dynamics, the system and bath Heisenberg operators satisfy the following equations:
\begin{eqnarray}\label{H-equations_TR}
\dot{Q_\mu}(t)=-\frac{\rmi}{\hbar}\left[H,Q_\mu(t)\right] \ \ , \ \ \dot{P_\mu}(t)=-\frac{\rmi}{\hbar}\left[H,P_\mu(t)\right],
\end{eqnarray}
where one has a change of sign respect to the usual Heisenberg dynamics (\ref{H-equations}).
By proceeding in the same way as in the usual case, we find the following modifications involving the matrix QLE satisfied by the central oscillator (\ref{Lang-Matr}):
\begin{eqnarray}
 &&{\bm H}(t)\rightarrow{\tilde{\bm H}}(t)=\left[\begin{array}{c c}
0 & \delta(t)\\
-\Omega_0^2\delta(t) & K(t)
\end{array}\right],\\
&&\eta(t)\rightarrow\tilde{\eta}(t)=\sum_i^N\epsilon_i\left[-Q_i(0)\cos\omega_it+\frac{P_i(0)}{\omega_i}\sin\omega_it\right].
\end{eqnarray}
This imply that the evolution is given as in equation (\ref{Lang_sol}), but with a change in the off-diagonal elements of the matrix propagator (\ref{M_prop}):
\begin{equation}
 {\bm \Phi}(t)\rightarrow{\tilde{\bm \Phi}}(t)=\left[\begin{array}{c c}
\dot{A}(t) & -A(t)\\
-\ddot{A}(t) & \dot{A}(t)
\end{array}\right],
\end{equation}
while the noise kernel (\ref{noise_kern}) remains unchanged.
By considering the above modifications, one straightforwardly find from (\ref{Wig_me}) the FP-like equation satisfied by the reduced Wigner for the central oscillator for the adjoint dynamics: it remains exactly the same except for a change in the signs of the drift term, the harmonic forcing term and the anomalous diffusion coefficient.
This imply that the limits generally considered in section (\ref{therm_limit}) for the usual process do also apply for the adjoint, moreover the late time density matrix is the same as in (\ref{Late-cov-mat}).

%%%%%%%%%%%%%%%%%%%%%%%%%%%%%%%%%%%%%%%%%%%%%%%%%%%%%%%%%
\section{Interaction energy term}\label{A-Int-term}

A way to express the average of $H_{\mathrm{I}}$, which is useful in the continuum limit, is the following. By plugging the Heisenberg formal solutions for the bath operators as functions of $Q(t)$ (\ref{Qi_t_formal}) in the interaction term, one gets
\begin{eqnarray}\label{H_int_t}
\hspace{-1cm} \average{H_\mathrm{I}(t)} =\average{\sum_i Q(t) Q_i(t)}=\average{Q(t)\eta(t)}+\average{\int_0^t\rmd s\;\dot{K}(t-s)Q(t)Q(s)},
\end{eqnarray}
where $\eta(t)$ is the fluctuating force term defined in (\ref{eta}).%\marginpar{Ricontrollare i segni!}
Using (\ref{Lang_sol}) and the initial absence of system-bath correlation, which implies $\average{Q(0)\eta(t)}=0$, one gets 
\begin{eqnarray}\label{Qeta}
\average{Q(t)\eta(t)} &=&-\int_0^t\rmd t'\;A(t-t')\average{\eta(t')\eta(t)}\nonumber\\
&=&-\int_0^\infty \rmd\omega\; \frac{\gamma(\omega)}{\omega^2}E(\omega,T)\int_0^t\rmd s\; A(s)\cos(\omega s).
\end{eqnarray}
Interestingly the use of the Ullerma strength (\ref{U_strenght}), from which one obtains that the damping coefficients are time-independent in the large cut-off limit, implies exactly the same integral form for the anomalous diffusion coefficient, so that $\average{Q(t)\eta(t)}=D_{qp}(t)$.  

In fact in the case of the Ullersma strength (\ref{U_strenght}) with large cut-off the diffusion matrix (\ref{eq:diffusion}) can be approximated by~\cite{Fleming}
\begin{equation}
 {\bm D}(t)=\frac{1}{2}\int_0^t\rmd\tau\;\left[{\bm \nu}(t-\tau){\bm \Phi} ^{\mathsf{T}}(t-\tau)+{\bm \Phi}(t-\tau){\bm \nu} ^{\mathsf{T}}(t-\tau)\right].
\end{equation}
This straightforwardly leads to the equivalence between $\average{Q(t)\eta(t)}$, which is part of the average interaction term (\ref{H_I_t}), and the anomalous diffusion term $D_{qp}(t)$.

The integral (\ref{Qeta}) is done by using the local propagator $A_{\loc}(t)$ (\ref{A_local}), by first integrating over time and then in the complex $\omega$ plane. One obtains
\begin{eqnarray}\fl\label{Dqp_t}
 &&D_{qp}(t) =\average{P^2}_{\eq}-\Omega_0^2\average{Q^2}_{\eq}-\left\{\dot{A}_{\loc}(t)+A_{\loc}(t)\left(2\Gamma-\frac{d}{dt}\right)\right\}F_C(t);\\ \fl
 &&F_C(t) =-(\ddot{C}(t)+\Omega_0^2C(t)+2\Gamma\dot{C}(t)).
\end{eqnarray}
The time dependent term contained in (\ref{Dqp_t}) vanishes so that one recovers the late-time anomalous diffusion coefficient (\ref{Dqp}), which is a positive quantity.
Here we have made use of the approximate equilibrium correlation function (\ref{C_l}). It follows that 
\begin{eqnarray}\fl
 F_C(t)=\kappa\hbar\left[-\frac{1}{2}\frac{\cot(\pi\alpha\tau_\beta) \rme^{-\alpha t}}{(1+({\Gamma}/{\alpha}))^2+({\Omega}/{\alpha})^2}+\frac{1}{\pi}\sum_{\ell=1}^\infty\frac{(\alpha\tau_\beta)^2}{(\alpha\tau_\beta)^2-\ell^2}\frac{\ell\rme^{-\ell t/\tau_\beta}}{(\ell+\Gamma\tau_\beta)^2+(\Omega\tau_\beta)^2}\right],
\end{eqnarray}
which can be approximated for $t\gg1/\alpha$ by
\begin{equation}
 F_C(t)\simeq\frac{\kappa\hbar}{\pi}\sum_{\ell=1}^\infty\frac{\ell\rme^{-\ell t/\tau_\beta}}{(\ell+\Gamma\tau_\beta)^2+(\Omega\tau_\beta)^2}.
\end{equation}
In the classical limit the anomalous diffusion coefficient vanishes.

To evaluate the second term of the sum in (\ref{H_int_t}), we can use the fact that  for large cut-off $\dot{K}(t)\sim-\kappa\alpha\delta(t)$.
Then one gets for the interaction term complete expression (\ref{H_I_t}).

In the finite case one has, by using the solutions (\ref{H-solutions}) and the initial conditions (\ref{factor_ic}):
\begin{eqnarray}\label{H_I_av_N}\fl
 \average{\sum_{i=1}^NQ(t)Q_i(t)} =\sum_{i=1}^N\left\{\epsilon_i\dot{A}_{00}\dot{A}_{i0}\average{Q^2(0)}+A_{00}A_{i0}\average{P^2(0)}+(\dot{A}_{00}A_{i0}+\dot{A}_{i0}A_{00})C_{qp}(0)\right.\nonumber\\
\qquad\left.{}+\sum_{j=1}^N\left(\frac{\dot{A}_{0i}\dot{A}_{ji}}{\omega_i^2}+A_{0i}A_{ji}\right)E(\omega_i,T)\right\}.
\end{eqnarray}
This expression is useful for a numerical calculation in the finite case.
By using the expression of the $A_{i0}(t)$ and the $A_{ij}(t)$ in function of the propagator $A(t)$, one obtains the expressions (\ref{H_int_t},\ref{Qeta}) exploited in the continuum limit.

%%%%%%%%%%%%%%%%%%%%%%%%%%%%%%%%%%%%%%%%%%%%%%%%%%%%%%%%%%%%%%%%%%%%%
\section{Evaluation of the bath entropy}\label{sec:bathentcalc}

One can straightforwardly evaluate the general quantum bath entropy (\ref{BathEnt}) if one finds a coordinate transformation that puts the density operator $\rho_\B$ in a normal form, namely a product of independent oscillator thermal states:
\begin{equation}\label{norm-bd}
\rho_\B=\bigotimes_\ell(1-\rme^{-\beta_\ell})\,\rme^{-\beta_\ell {n}_\ell},
\end{equation}
where ${n}_\ell=a^\dagger_\ell a_\ell$, with ${a}_\ell=( {q}_\ell+\rmi {p}_\ell)/\sqrt{2}$, and where the $\beta_\ell$ are suitable effective inverse temperatures. In fact, by putting the density operator in this form, the calculation of entropy is easily obtained by carrying the trace over the space of the eigenstates of the number operator: $\left| {n}_1, {n}_2,\ldots, {n}_\ell,\ldots, {n}_N\right\rangle$. One obtains
\begin{equation}\label{b-entropy}
S_{\B}= \sum_\ell \big( (k_\ell+1/2)\ln(k_\ell+1/2)-(k_\ell-1/2)\ln(k_\ell-1/2) \big),
\end{equation}
where $k_\ell=\frac{1}{2}\coth\left(\frac{1}{2}\beta_\ell\right)=\average{ {q}_\ell^2}=\average{ {p}_\ell^2}=\average{ {n}_\ell}+\frac{1}{2}$. For simplicity we have put here $\hbar=1$.

We know from (\ref{ME}) that the reduced density matrix for the bath
is Gaussian. Here first moments can be shifted to $0$, as this transformation leaves the entropy invariant.
Then from an informational point of view the bath is fully characterized by the covariance matrix $\sigma_{ij}^\B$ (\ref{sigma_bath}). 

The normal form (\ref{norm-bd}) and values of the $k_\ell$'s can be actually recovered by a ``pseudo-diagonalization'' of the correlation matrix. 
This can be done using a symplectic transformation, $\xi \mapsto S \xi$ where $S$ is a $2N\times2N$-matrix, i.e. a transformation preserving the bosonic commutation rules:
\begin{equation}\label{sympl-T}
\bm{\beta}=S \bm{\beta} S ^{\mathsf{T}},
\end{equation}
where
\begin{equation}
\bm{\beta}=\left(\begin{array}{cc}
        0 & \bm{1} \\
	-\bm{1} & 0
       \end{array}
	\right);\qquad \bm{1}=(\delta_{k,\ell}),\quad k,\ell=1,2,\ldots,N.
\end{equation}
One then choose $S$ such that the correlation matrix in the new basis is diagonal:
\begin{equation}
 \sigma^\B\mapsto{\sigma'}^{\B}=S\sigma^BS ^{\mathsf{T}}=\mathop{\mathrm{diag}}(\kappa_1,\kappa_2,\ldots,\kappa_N,\kappa_1,\kappa_2,\ldots,\kappa_N).
\end{equation}
This can always be done, as affirmed by Williamson's theorem \cite{Williamson}, due to the fact that the correlation matrix is symmetric and  positive definite. Due to the particular block form of the correlation matrix, the $k_\ell$'s are doubly degenerate, as shown in~\cite{Colpa}. 

The pseudo-eigenvalues and the symplectic matrix $S$ can be obtained, as explained in~\cite{Simon}, by diagonalizing the symmetric matrix $K\beta \sigma^\B\beta ^{\mathsf{T}}K ^{\mathsf{T}}$, where the matrix $K$ is obtained by a Cholesky decomposition of the correlation matrix:
\begin{equation}
 \sigma^B=K ^{\mathsf{T}}K.
\end{equation}
This can actually be carried out, since $\sigma^B$ is positive definite. The eigenvalues one finds are actually the doubly-degenerate squares of the pseudo-eigenvalues $k_\ell$.

Given an operator $\hat{A}$, its Wigner transform is defined by~\cite{Hillery}
\begin{equation}
A(p, q)=\int \rmd z\;\rme^{\rmi pz/\hbar}\,\left<q-\frac{z}{2}\right|\hat{A}\left|q+\frac{z}{2}\right>.
\end{equation}
In the general quantum case, the Wigner transform of $\ln \rho_\B$ is $-\xi ^{\mathsf{T}}M\xi-\ln Z_\B$, since one has
\begin{equation}
\rho_\B=\exp{\{- {\xi} ^{\mathsf{T}}M {\xi}\}}/Z_\B,
\end{equation}
where $M$ is a square $2N\times2N$ matrix. This matrix transforms under a symplectic transformation of the phase space operators (\ref{sympl-T}) like $(\sigma^\B)^{-1}$:
\begin{equation}
M \mapsto M'=(S ^{\mathsf{T}})^{-1}M S^{-1}.
\end{equation}
In the classical limit the diagonalized matrices $M'$ and $({\sigma'}^\B)^{-1}$, coincide and therefore also $(\sigma^B)^{-1}$ and $M$ have to coincide. Therefore in the classical limit the Wigner distribution corresponding to the density operator $\rho_\B$ has the same expression as the classical probability distribution apart from multiplicative coefficients, i.e.,
\begin{equation}\label{WB_CL}
W_\B(q,p,t)= \exp{\{-\frac{1}{2} {\xi^\dagger} (\sigma^\B)^{-1} {\xi}\}} /\left[(2\pi)^N(\det \sigma^\B)^{1/2}\right].
\end{equation}
where $\sigma^\B$ is given in eq.~(\ref{sigma_bath}). Then the entropy of the bath can be easily calculated via a Gaussian integral:
\begin{eqnarray}\label{SBCLcov}
S_\B &=&-\int \rmd q\,\rmd p\; W_\B(q,p,t)_\mathrm{CL}\ln \left[(2\pi)^N W_\B(q,p,t)_\mathrm{CL}\right]\\
&= &N + \ln (\det \sigma^\B)^{1/2} \nonumber.
\end{eqnarray}
This result can be also be obtained by noticing that in the classical
limit, where $k_\ell\gg \hbar/2$, the expression (\ref{b-entropy}) for the bath entropy reduces to
\begin{equation}
S_\B=N+\ln\prod_{\ell=1}^N k_\ell = N + \ln (\det S\sigma^BS ^{\mathsf{T}})^{1/2} .
\end{equation}
Thus, since $\det S=1$ (\ref{sympl-T}) we recover (\ref{SBCLcov}).

%%%%%%%%%%%%%%%%%%%%%%%%%%%%%%%%%%%%%%%%%%%%%%%%%%%%%%%%%%%%%%%%%%%%
\section{Bath covariance matrix}\label{app:eqtime}

The covariance matrix of the bath is defined as:
\begin{equation}\label{sigma_bath}
\sigma_{ij}^\B =\frac{1}{2}\langle\{ {\xi}_i, {\xi}_j\}\rangle-\langle {\xi}_i\rangle\langle {\xi}_j\rangle,
\end{equation}
where ${\xi} =(Q_1,...,Q_N,P_1,...,P_N)$, $i$ and $j$ identify the bath oscillators.

Using the Heisenberg solutions (\ref{H-solutions}), the variance of a bath position operator with average taken over the initial conditions (\ref{factor_ic}) give, with help of (\ref{bath_ic}):
\begin{eqnarray}\label{var_bath_t}
&&\average{Q_i^{2}(t)}-\average{Q_i(t)}^2=\dot{A}_{i0}^{2}\sigma_q^2(0)+A_{i0}^2\sigma_p^2(0)+2\dot{A}_{i0}A_{i0}C_{qp}(0)\nonumber\\
&&\qquad\qquad\qquad\qquad+\sum_{\ell=1}^{N}\left[\frac{\dot{A}_{i\ell}^2}{\omega_\ell^2}+A_{i\ell}^2\right]E(\omega_\ell,T) .
\end{eqnarray}
The last sum can be rewritten, using expression (\ref{A_ij(A_i0)}) for the $A_{i\ell}$'s, as follows:
\begin{eqnarray}\fl
&&\sum_{\ell=1}^{N}\left[\frac{\dot{A}_{i\ell}^2}{\omega_\ell^2}+A_{i\ell}^2\right]E(\omega_\ell,T)=\sum_{\ell=1}^N\frac{\epsilon_\ell^2}{\omega_\ell^2}\left|\int_0^t\rmd\tau\; A_{i0}(\tau)e^{\rmi\omega_\ell\tau}\right|^2E(\omega_\ell,T)\nonumber\\
&&\qquad\qquad\qquad+\frac{E(\omega_i,T)}{\omega_i^2}+\frac{2\epsilon_i}{\omega_i^2}\int_0^t\rmd\tau\;A_{i0}(\tau)\cos(\omega_i\tau)E(\omega_i,T).
\end{eqnarray}
It contains a term explicitly depending on the initial conditions of the central oscillator, and a thermal part. The latter one is made of a term explicitly depending on the initial conditions of the bath oscillator, a sum of the kind $\sum_{\ell=1}\epsilon_\ell^2\ldots$, which is easily put into integral form by using the strength (\ref{gamma_omega}), plus an integral containing $A_{i0}(t)$ and an oscillating function of time, multiplied by the coupling $\epsilon_i$ of the oscillator. The same structure is obtained for  every term of the bath covariance matrix $\sigma_{ij}^\B$ between any momentum and bath operators.

%%%%%%%%%%%%%%%%%%%%%%%%%%%%%%%%%%%%%%%%%%%%%%%%%%%%%%%%%%%%%%%%%%%%%%
\section*{References}

%\bibliographystyle{iopart-num}
%\bibliography{quantumBrownian}

\providecommand{\newblock}{}

%%%%%%%%%%%%%%%%%%%%%%%%%%%%%%%%%%%%%%%%%%%%%%%%%%%%%%%%%%%%%%%%%%%%%%

\end{document}